\documentclass{knac}

\usepackage{hyperref}
\usepackage{booktabs}
\usepackage{fancyvrb}
\usepackage{algorithm}
\usepackage{algorithmic}
\usepackage{hypcap}
\usepackage{multirow}
\let\captionbox\relax
\usepackage{subcaption}

\usepackage{natbib}   
\setcitestyle{numbers,square,comma}
\makeatletter
\renewcommand{\@biblabel}[1]{#1.\hspace{1 em}}

\makeatother

\begin{document}
\title{AI Agents as Policymakers in Simulated Epidemics}


\author{%
Goshi Aoki$^{1}$ and Navid Ghaffarzadegan$^{1,+}$\\[1ex]
$^{1}$Department of Industrial and Systems Engineering, Virginia Tech, Alexandria, VA, USA\\[0.8ex]
$^{+}$Corresponding author: navidg@vt.edu
}


\begin{abstract}
AI agents are increasingly deployed as quasi-autonomous systems for specialized tasks, yet their potential as computational models of decision-making remains underexplored. We develop a generative AI agent to study repetitive policy decisions during an epidemic, embedding the agent—prompted to act as a city mayor—within a simulated SEIR environment. Each week, the agent receives updated epidemiological information, evaluates the evolving situation, and sets business restriction levels. The agent is equipped with a dynamic memory that weights past events by recency and is evaluated in both single- and ensemble-agent settings across environments of varying complexity. Across scenarios, the agent exhibits human-like reactive behavior, tightening restrictions in response to rising cases and relaxing them as risk declines. Crucially, providing the agent with brief systems-level knowledge of epidemic dynamics—highlighting feedbacks between disease spread and behavioral responses—substantially improves decision quality and stability. The results illustrate how theory-informed prompting can shape emergent policy behavior in AI agents. These findings demonstrate that generative AI agents, when situated in structured environments and guided by minimal domain theory, can serve as powerful computational models for studying decision-making and policy design in complex social systems.

\end{abstract}



\section{Introduction}

An AI agent is a quasi-autonomous computational entity that perceives information from its environment, maintains internal state or memory, and makes decisions aimed at achieving specific goals \citep{Wooldridge_Jennings_1995}. Applications span a range from simulated games \citep{Park2023GenerativeAgents} to service industries, including customer support, virtual assistants, education, and entertainment \citep{chong_ai-chatbots_2021, knani_artificial_2022}. Such agents often leverage large language models (LLMs) and can be trained with background knowledge, personalities, and memory structures, enabling them to process information, adapt their behavior over time, and interact with their environment in a human-like, goal-directed manner \citep{Park2023GenerativeAgents, xu_ai_2024}. By combining memory, reasoning, and action, these agents bridge the gap between traditional AI systems and more realistic, human-centric simulations \citep{Park2023GenerativeAgents}.

Interest in generative AI agents has grown rapidly due to their potential to model complex social interactions, support decision-making, and enhance virtual environments \citep{xi_rise_2025, wang_survey_2024, gao_large_2024, bail2024can}. Their ability to learn from context, remember past experiences, and exhibit individualized behavior makes them valuable for research in social science, education, gaming, and policy simulation \citep{xu_ai_2024, bail2024can, ziems_can_2024}. Recent work on generative AI agents combines LLMs with memory, reflection, and planning mechanisms to support coherent long-term behavior. Park et al. introduced an architecture in which agents maintain a stream of experiences, retrieve relevant memories, and synthesize reflections to guide actions in an interactive environment \citep{Park2023GenerativeAgents}. In their small-town simulation, agents exhibit believable daily routines and emergent social behaviors from minimal initial instructions, though the authors emphasize that such behavior reflects believability rather than genuine human-like agency \citep{Park2023GenerativeAgents}. This work builds on earlier "social simulacra" methods that use LLMs to instantiate plausible online communities for design prototyping \citep{Park2022SocialSimulacra}.

A key challenge is validating these agent societies beyond anecdotal plausibility. Park et al. address this by scaling to over a thousand generative personas and testing whether simulated populations reproduce empirical regularities in survey-style responses \citep{Park2024ThousandPeople}. Similarly, Argyle et al. show that carefully conditioned LLM outputs can approximate distributions of human survey responses \citep{Argyle2023OutOfOneMany}. Giabbanelli et al. enhance methodological rigor by systematically evaluating LLM-based approaches for converting agent-based modeling (ABM) outputs into empathetic narratives \citep{Giabbanelli03092025}. These studies shift the field from demonstration toward systematic validation against external benchmarks.

This trajectory is reflected in generative ABM (GABM). Ghaffarzadegan et al. describe GABM as coupling mechanistic models with LLM-powered decision-making, allowing decision rules to emerge from prompts rather than fixed equations \citep{Ghaffarzadegan2024GABM}. Their tutorial shows that GABM outcomes can be sensitive to prompt wording and ordering, motivating transparent prompt reporting and sensitivity analysis.

Williams et al. apply this approach to epidemic modeling \citep{Williams2023EpidemicGABM}. Their agents receive symptom and prevalence information and adjust mobility accordingly; the resulting population dynamics reproduce risk-responsive behavior and multi-wave trajectories. Other studies examine LLM behavior in strategic settings more directly. Lorè and Heydari test LLMs across social dilemmas, finding that contextual framing significantly influences strategic choices, which shows high sensitivity to context but limited capacity for abstract strategic reasoning \citep{Lore2024Strategic}. Schmidt et al. probe GPT's advice in Dictator and Ultimatum Games, finding that the model captures reciprocity and fairness but fails to adjust for strategic risk: suggested offers in the Dictator Game exceeded those in the Ultimatum Game, reversing the pattern observed in human behavior \citep{Schmidt2024GPT35}. These findings highlight both the promise and limitations of LLM-based agents in strategic contexts. Additional work demonstrates LLM capabilities in repeated games \citep{Akata2023RepeatedGames} and social network diffusion \citep{Gao2023SocialNetwork}. 

Emerging applications of generative agents involve more complex and sensitive tasks that require decision-making \citep{xu_ai_2020, meta2022human}. A particularly promising area is policymaking. The question is "Can agents move beyond analytical support to actively inform or even make repetitive policy decisions?" Such contexts are high-stakes and impactful, demanding careful exploration. The extent to which generative agents can act as high-level decision makers, and methods to improve their performance, remain open questions. The current study moves in this direction, investigating the potential of generative AI agents to operate in policy environments. 


Since testing policy agents in the real world is not yet feasible, we aim to test their performance in a simulated world. Specifically, in this paper, we develop a policymaker AI agent that continuously receives information and makes repeated decisions. In this context, the agent sets business closures and restrictions that influence societal interactions and disease spread. This specific decision-making case is used due to the relevance of repetitive decision-making and the fact we could manipulate the level of complexity by including human behavior adaptation \citep{espinoza_adaptive_2024, pant_mathematical_2024, fenichel_adaptive_2011}. Nevertheless, the concept can be generalized to similar repetitive decision-making contexts. In addition, simulation models of epidemics are well-established \citep{kermack_contribution_1927, ross_application_1917, ferguson_capturing_2007}, letting us focus the modeling effort on the agent-side.

\section{Methods}

\subsection{Conceptual Model}

\begin{figure}[t]
    \centering
    \includegraphics[width=0.75\textwidth]{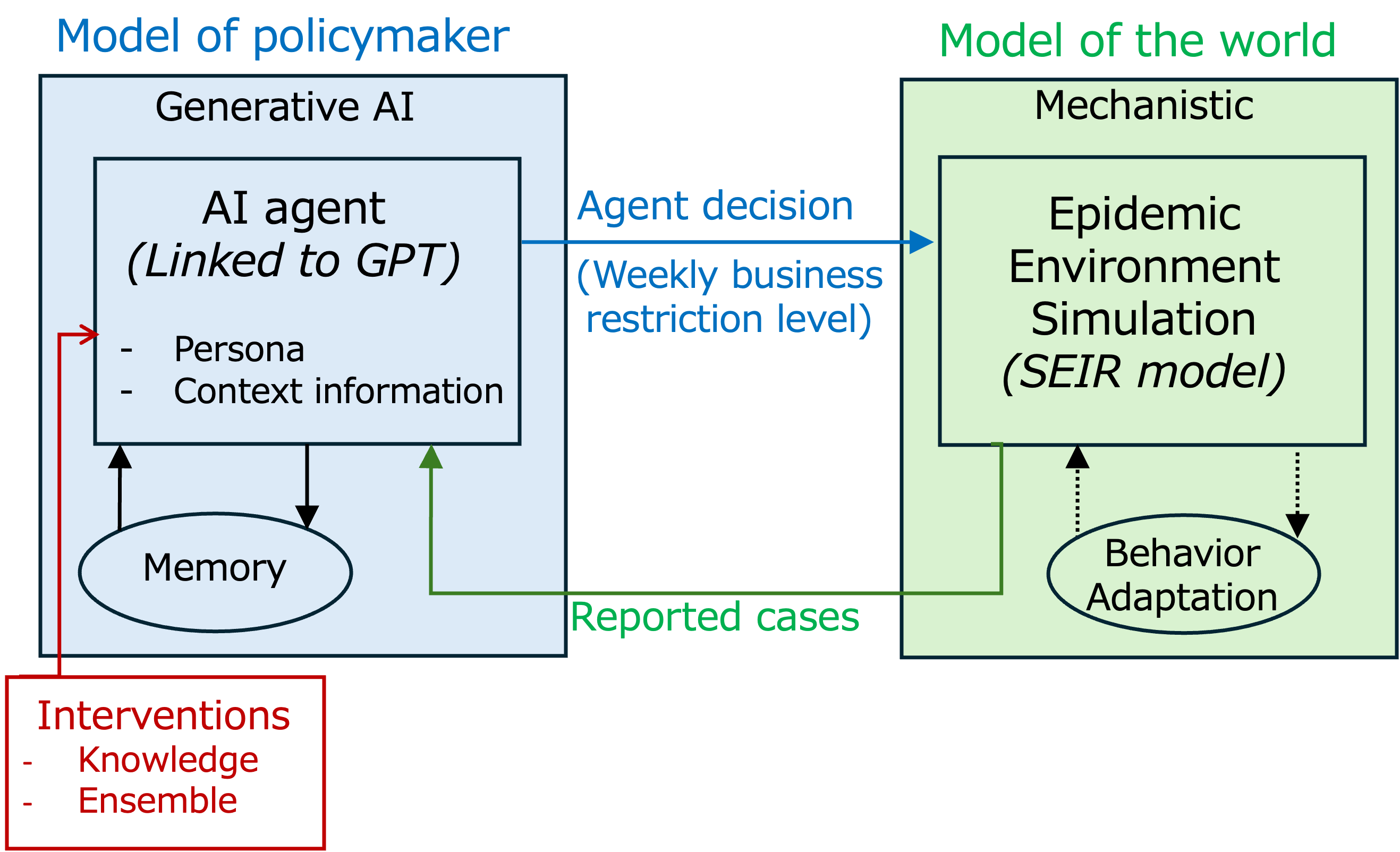}
    \caption{Conceptual architecture of an AI agent as a policymaker (left) coupled with a mechanistic model of simulated epidemic (right). 
    Note: Every week the policymaking agent observes recent cases and retrieved memories, outputs a business-restriction level; the SEIR environment advances affected by the policy, internal dynamics of the spread of virus, and in one of the world-scenarios with public behavioral response (dotted lines).}
    \label{fig:concept}
\end{figure}

Figure~\ref{fig:concept} illustrates the overall architecture. As stated, we develop a policymaker AI agent that makes repeated decisions within a simulated environment. The context resembles policymaking during the COVID-19 pandemic, where policymakers could influence societal interactions and disease spread by regulating business activities—for example, closing restaurants, bars, or schools, or, in extreme situations, implementing a complete lockdown. To simulate the epidemic environment, we use a simple mechanistic model of disease spread, commonly referred to as SEIR. 

Every seven days, the AI agent receives information about the state of the disease and, using its memory and background knowledge, forecasts the next week’s trajectory, reasons about potential outcomes, and decides the level of business restrictions. To support this reasoning and decision-making, the agent is coupled with an LLM (here, GPT-5 nano \citep{OpenAI2025GPT5}), enabling it to act quasi-autonomously without direct intervention from the modeler. Once a decision is made, it is fed to the simulated environment and implemented, after which the model runs for a 7-day period. This process continues for a year (365 days).

We examine the AI agent's forecasts and decisions in a 2 (world model) × 4 (AI agent) experimental design. Specifically, our experiments consider two major scenarios of the epidemic environment. In the first, the disease spreads according to a standard SEIR model, coupled with continuous policy decisions. In the second, we employ a modified SEIR model, often referred to as SEIRb (with “b” for behavior), in which public responses to perceived risk are incorporated into the mechanistic model. This scenario, termed “behavioral adaptation,” provides a more realistic context in which both policies and voluntary public behaviors influence disease dynamics. In both scenarios, the agent is informed about the state of the disease, and its policy decisions feed back into the epidemic model, with stricter restrictions slowing disease spread. In the behavioral adaptation scenario, voluntary public responses to risk further increase the complexity of the world model.

To further examine the agent’s performance, in addition to the base run, we implemented three major interventions: (1) providing systems-level knowledge about epidemic feedback loops in a short textual format, (2) using an ensemble of policy agents with aggregated decisions, and (3) combining both interventions.

Altogether, this results in a 2 (environment: simple vs. behavioral adaptation) × 4 (agent: base, knowledge, ensemble, ensemble with knowledge) experimental design. Each configuration is repeated for 10 independent simulation runs (with the same random seeds) to assess variability. Simulations run for 365 days, with the agent beginning active decision-making from Week 6 to allow initial warm-up period and memory construction. Across these experiments, we evaluate the agent’s forecasting accuracy, policy decisions, and epidemic outcomes in terms of cases.

\subsection{Model of the World: Epidemic Environment Simulation}

We consider two alternative scenarios for the epidemic environment. In both cases, the core epidemic dynamics are formulated using a conventional Susceptible--Exposed--Infectious--Recovered (SEIR) compartmental model. Let $S(t)$, $E(t)$, $I(t)$, and $R(t)$ denote the number of individuals in each compartment at time $t$ (day), with total population
$N = S(t) + E(t) + I(t) + R(t)$. The continuous-time dynamics are given by:
\begin{align}
\frac{dS}{dt} &= -\beta \frac{S I}{N}, \\
\frac{dE}{dt} &= \beta \frac{S I}{N} - \frac{E}{L}, \\
\frac{dI}{dt} &= \frac{E}{L} - \frac{I}{D}, \\
\frac{dR}{dt} &= \frac{I}{D},
\end{align}
where $\beta$ is the transmission rate, $L$ is the latent period, and $D$ is the infectious period.

To represent the effects of voluntary public behavior, government policy interventions, and stochastic variation, we decompose the transmission rate $\beta$ as
\begin{align}
\beta &= \beta_0 \, b \, g \, \epsilon ,
\end{align}
where $\beta_0$ is the baseline infectivity in the absence of government and public responses,
$b$ is a behavioral modifier capturing voluntary public response,
$g$ represents the effect of government restrictions ($G$),
and $\epsilon \sim \text{Uniform}(0.5, 1.5)$ denotes daily stochastic noise.

The two epidemic scenarios differ in how voluntary behavioral adaptation is modeled. To model behavioral responses, we adopt a previously validated framework referred to as SEIR\textit{b}, where \textit{b} denotes behavior, which captures feedback between epidemic dynamics and behavior in a simple algebraic form \citep{rahmandad_enhancing_2022, lejeune_mathematical_2024, lejeune_formulating_2025}. Specifically, in the first (simpler) scenario, no behavioral adaptation is assumed and $b = 1$. In the second scenario, behavioral adaptation is explicitly modeled as a function of perceived risk, proxied by lagged reported cases $C_{t-1}$. The effects of government restrictions ($0 \leq G \leq 1$) and perceived risk on infectivity are specified as follows:
\begin{align}
g &= 1 - \alpha G, \\
b &=
\begin{cases}
1, & \text{World 1 (no behavioral adaptation)}, \\
\displaystyle \frac{1}{1 + k C_{t-1}}, & \text{World 2 (behavioral adaptation)}.
\end{cases}
\end{align}

Time is discretized in daily steps using numerical integration. Daily reported cases are defined as the flow from the exposed to the infectious compartment,
$C(t) = E(t) / L$. The weekly case count provided to the policy agent is computed as the mean of daily reported cases over the preceding seven days.

Table~\ref{tab:seir_params} summarizes the parameter values.

\begin{table}[htbp]
\centering
\caption{SEIR model parameters.}
\label{tab:seir_params}
\small
\begin{tabular}{lcc}
\toprule
Parameter & Symbol & Value \\
\midrule
Population size & $N$ & $10^6$ \\
Initial susceptible & $S_0$ & 999,999 \\
Initial exposed & $E_0$ & 0 \\
Initial infected & $I_0$ & 1 \\
Baseline transmission rate & $\beta_0$ & 0.2 day$^{-1}$ \\
Latent period & $L$ & 4 days \\
Infectious period & $D$ & 10 days \\
Government decision effect & $\alpha$ & 0.8 \\
Human behavior sensitivity & $k$ & $5 \times 10^{-4}$ \\
\bottomrule
\end{tabular}
\end{table}

\subsection{Model of Policymaker: Generative AI Agent}
The policymaker agent is implemented as a generative AI system prompted to role-play as Jennifer, a city mayor (see Appendix for the prompt template and a complete example). The agent's persona emphasizes evidence-based decision-making, transparency, and balancing public health with economic considerations.

Each week, the agent receives a structured prompt containing contextual information about her role as a democratically elected mayor of a city with a population of one million, guiding principles emphasizing the importance of both economic activity and public health, and a default stance favoring keeping the city open when feasible. The prompt also includes the current state of the epidemic, summarized by reported cases over the previous seven days, along with information about the policy implemented in the prior period. In addition, the agent retrieves information about historical epidemic trends from its memory.

The agent is prompted to (i) forecast next week’s cases assuming a full opening, (ii) select the level of business restrictions ($G$) to be implemented, (iii) forecast next week’s cases under the newly chosen policy, and (iv) provide a brief textual explanation justifying its decision. All outputs are returned in a structured JSON format.

\paragraph{Memory Architecture.}
\label{method:memory}
The memory architecture supports stochastic retrieval weighted by recency. Memories are stored as tuples of $(\text{week}, \text{restriction level}, \text{actual cases on the decision day})$. At each decision point, five past events—comprising historical case levels and policy decisions—are sampled randomly, with probabilities weighted by recency. Specifically, the retrieval weight for memory $i$ in a store of $n$ memories is defined as

\begin{equation}
w_i \propto \exp\!\bigl(0.1\, (i - n)\bigr),
\end{equation}
where $i = 0$ denotes the oldest memory and $i = n-1$ the most recent. We normalize the weights dividing them with the sum of the weights, so the total is equal to one. This weighting scheme favors recent experiences while preserving a nonzero probability of recalling earlier events. In this context, we treated all pieces of information equally important (assuming successful and failure experiences are equally important) and relevant (since all pieces of information in the memory are related to past policy decisions), thus the probability of retrieval is solely a function of recency. 

\paragraph{Interventions.}
In addition to the base run, the agent is evaluated under three intervention conditions. In the first intervention, referred to as the \textit{Knowledge} condition, an educational paragraph is appended to the prompt explaining epidemic dynamics through the lens of dynamic complexity and feedback loops \citep{bayer2004business}. This text is written in plain language, avoids technical jargon, and is tailored to match the active feedback structure in each epidemic scenario. Consequently, the content differs slightly across world models to accurately reflect their underlying dynamics. For example, in World 2 (behavioral adaptation) scenario, the knowledge text reads as follows:

\begin{quote}
\small
\textit{``Here I provide some information to help you better understand epidemic dynamics. Epidemics are primarily governed by interacting feedback loops. First, there is the reinforcing (positive) feedback loop: infection breeds more infection. Without intervention, cases grow exponentially as each infected person spreads the disease to susceptible individuals. The second is the balancing feedback loop of depletion. As more people become infected and then immune after recovery, the pool of susceptible individuals shrinks, which naturally slows transmission over time.
In addition to these biological feedback loops, there are behavioral feedback loops that shape transmission. As cases rise, people tend to grow more cautious and voluntarily adopt protective behaviors such as masking, distancing, and avoiding crowds. These reactions reduce the transmission rate. Conversely, when cases decline, individuals often relax their guard, which can lead to increased transmission and a resurgence of cases. Most importantly for your role, government restrictions are also part of a behavioral feedback loop. As you impose stricter measures on business and social activities, the probability of disease spread decreases. In simple terms, stronger shutdowns mean lower future infection rates—though naturally at an economic cost. Your shutdown decisions do not operate in isolation; they interact with voluntary citizen behavior driven by perceived risk.
When forecasting and making decisions, it is crucial to recognize that implementing or relaxing restrictive policies influences the spread of the disease and people's responses to those changes.''}
\end{quote}

In the second and third interventions, AI-agent ensembles are used. In these conditions, final policy decisions and forecasts are obtained by averaging outputs across ten independent agent instances. Operationally, the model performs ten independent API calls, and numeric outputs—including the level of business restrictions and both case forecasts—are averaged. 

\paragraph{Evaluations.}
We assess agent performance using three complementary approaches. First, we visualize daily case trajectories and weekly reductions in transmission over the 365-day simulation period. Second, we compare aggregate outcomes across conditions using three metrics: (i) cumulative cases, representing total infections over the simulation; (ii) cumulative prediction error, defined as the sum of absolute differences between the agent’s weekly case predictions and realized outcomes; and (iii) mean reduction in transmission. For each metric, we compute the mean across 10 independent simulation runs per condition, with error bars indicating the range (minimum to maximum) to capture variability arising from stochastic epidemic dynamics and LLM sampling. Third, we conduct a systematic analysis of the agent’s decisions using regression analyses to identify factors that correlate with the AI agent’s predictions and policy choices.

\section{Results}
\label{sec:results}

\subsection{World 1: Policy as the Primary Behavioral Lever}
\label{sec:results_world1}

\begin{figure*}[t]
    \centering
    \includegraphics[width=0.8\textwidth]{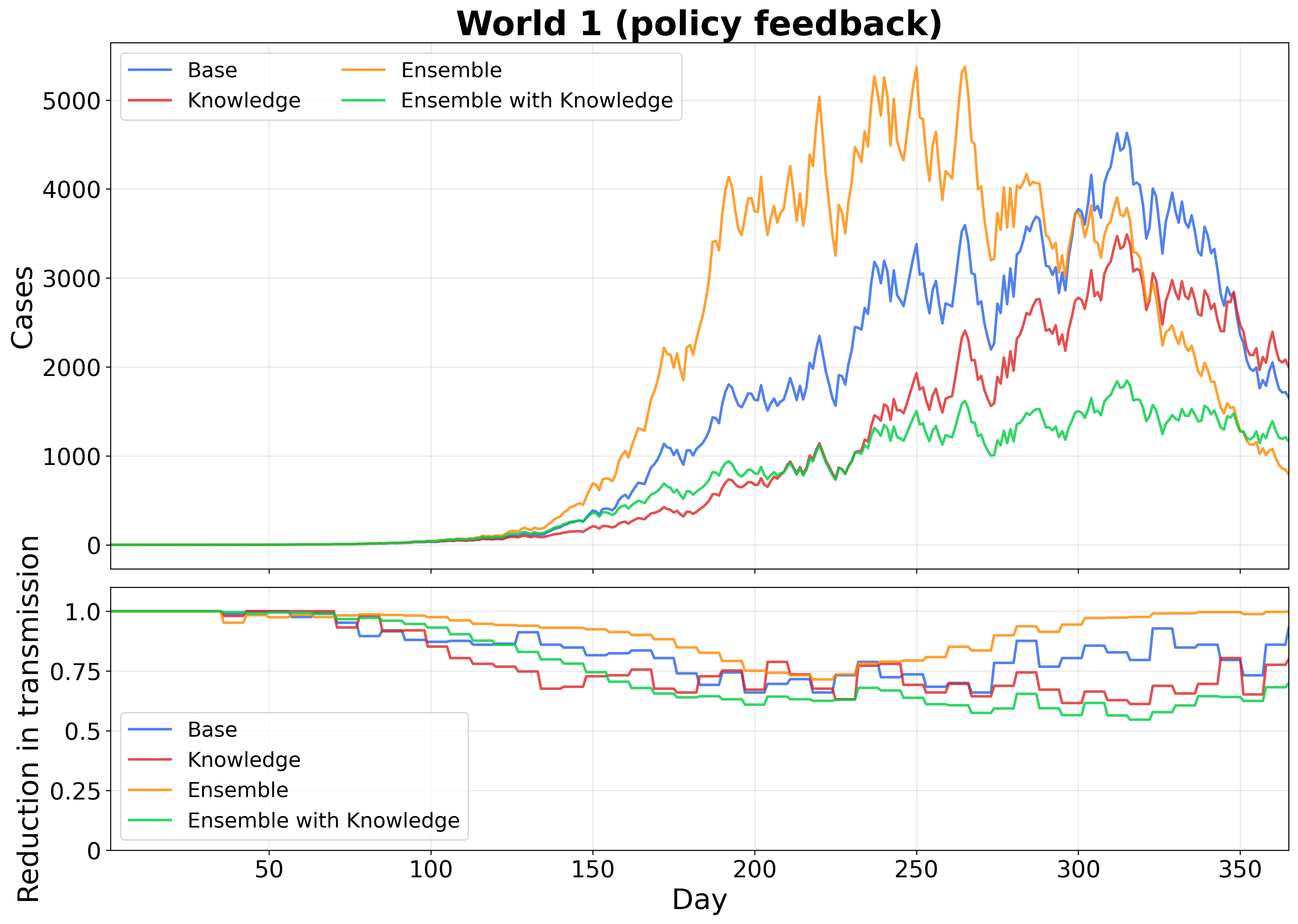}
    \caption{
    \textbf{World~1 (policy feedback):} Mean trajectories over 10 simulation runs.
    Top panel: reported cases; bottom panel: reduction in transmission implied by policy (lower = stronger suppression).
    }
    \label{fig:world1_timeseries}
\end{figure*}

Figure~\ref{fig:world1_timeseries} shows mean trajectories over 10 simulation runs, with reported cases in the top panel and reductions in transmission in the bottom panel. Across all four conditions, trajectories exhibit qualitatively similar dynamics: as cases rise, policy responses impose stronger restrictions, leading to lower transmission and, subsequently, fewer cases. The scenarios differ, however, in how quickly and consistently agents react to the epidemic, with some exhibiting earlier and stronger actions leading to an earlier decline in transmission.

In the \textit{Base} condition (blue), the AI agent exhibits largely reactive policymaking. The \textit{Knowledge} condition (red), in which the agent is provided with a brief description of epidemic feedback loops, follows a similar reactive pattern but intervenes earlier and more aggressively. As a result, the transmission rate declines sooner and remains lower for a longer period over the course of the outbreak. This earlier suppression keeps case counts below the \textit{Base} trajectory, reducing year-long cumulative cases by approximately one third.

The \textit{Ensemble without Knowledge} condition (yellow) underperforms the \textit{Base} case. Reductions in transmission remain weak compared with other configurations, indicating minimal policy intervention. This scenario consists of 10 distinct agents whose final action is determined by averaging their individual decisions. Due to stochasticity in LLM outputs, a single poor decision by one agent (e.g., removing all restrictions) can disproportionately influence the aggregate decision and resulting outcomes.

The \textit{Ensemble with Knowledge} configuration (green) achieves the strongest performance. Reductions in transmission emerge early and remain substantial, producing the flattest case trajectory and the lowest peak levels. This configuration yields the lowest mean year-long cumulative cases—approximately half of those observed in the \textit{Base} condition—suggesting that ensemble averaging is most effective when paired with causal knowledge about the epidemic environment.

\begin{figure*}[ht]
\centering
\begin{subfigure}[b]{0.32\textwidth}
    \includegraphics[width=\textwidth]{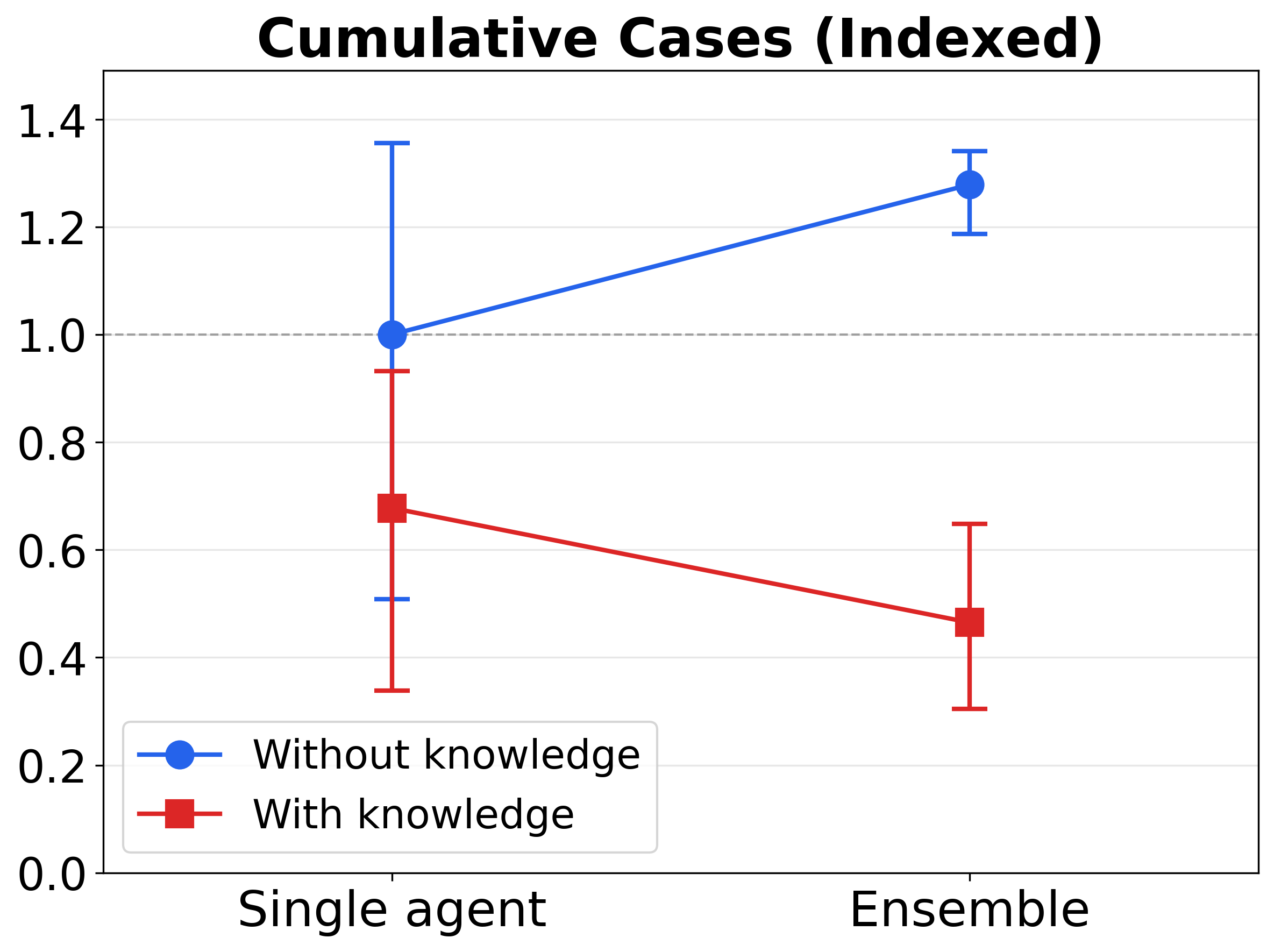}
    \caption{Cumulative cases}
    \label{fig:world1_cases}
\end{subfigure}
\hfill
\begin{subfigure}[b]{0.32\textwidth}
    \includegraphics[width=\textwidth]{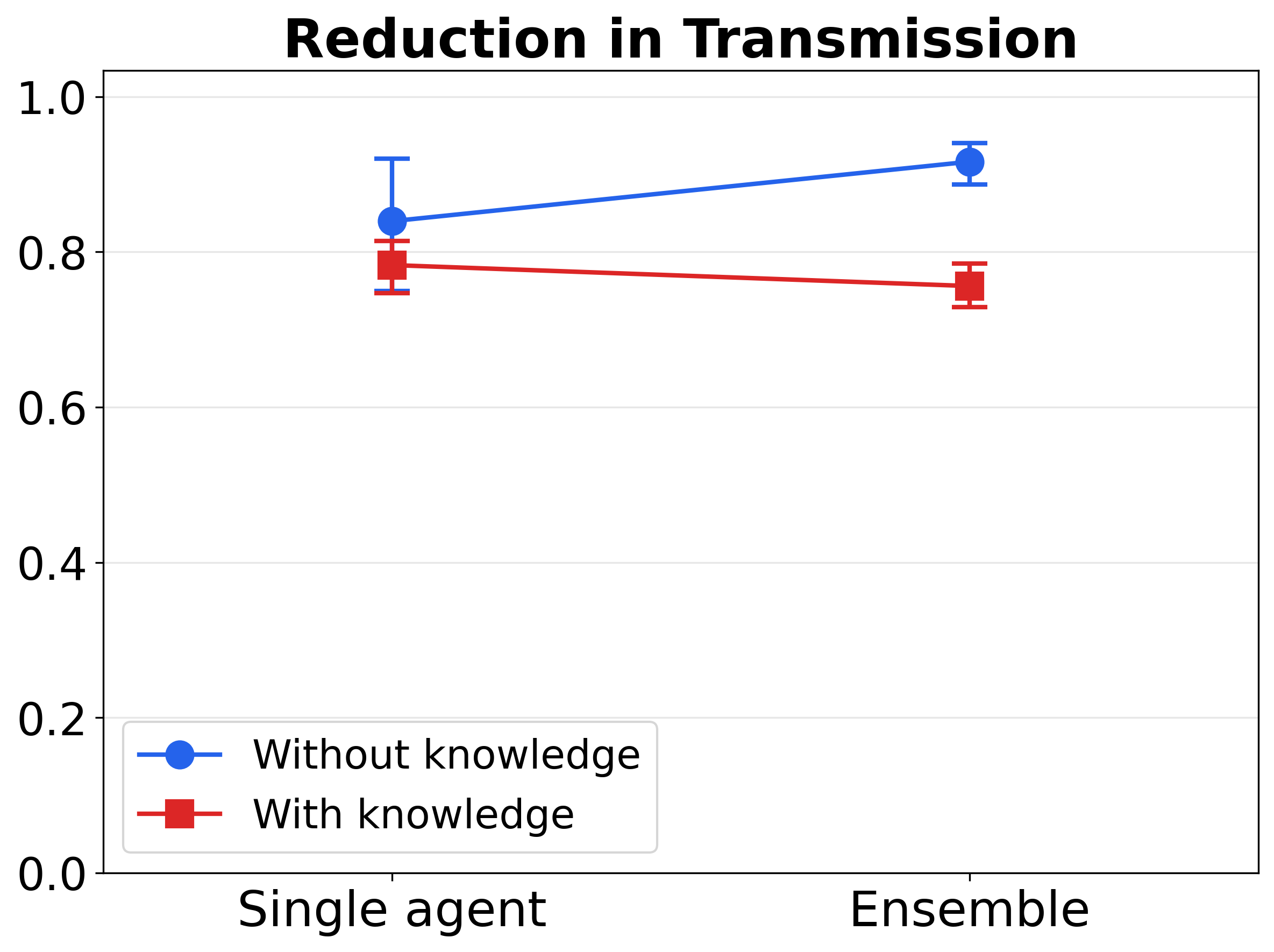}
    \caption{Reduction in transmission}
    \label{fig:world1_reduction}
\end{subfigure}
\hfill
\begin{subfigure}[b]{0.32\textwidth}
    \includegraphics[width=\textwidth]{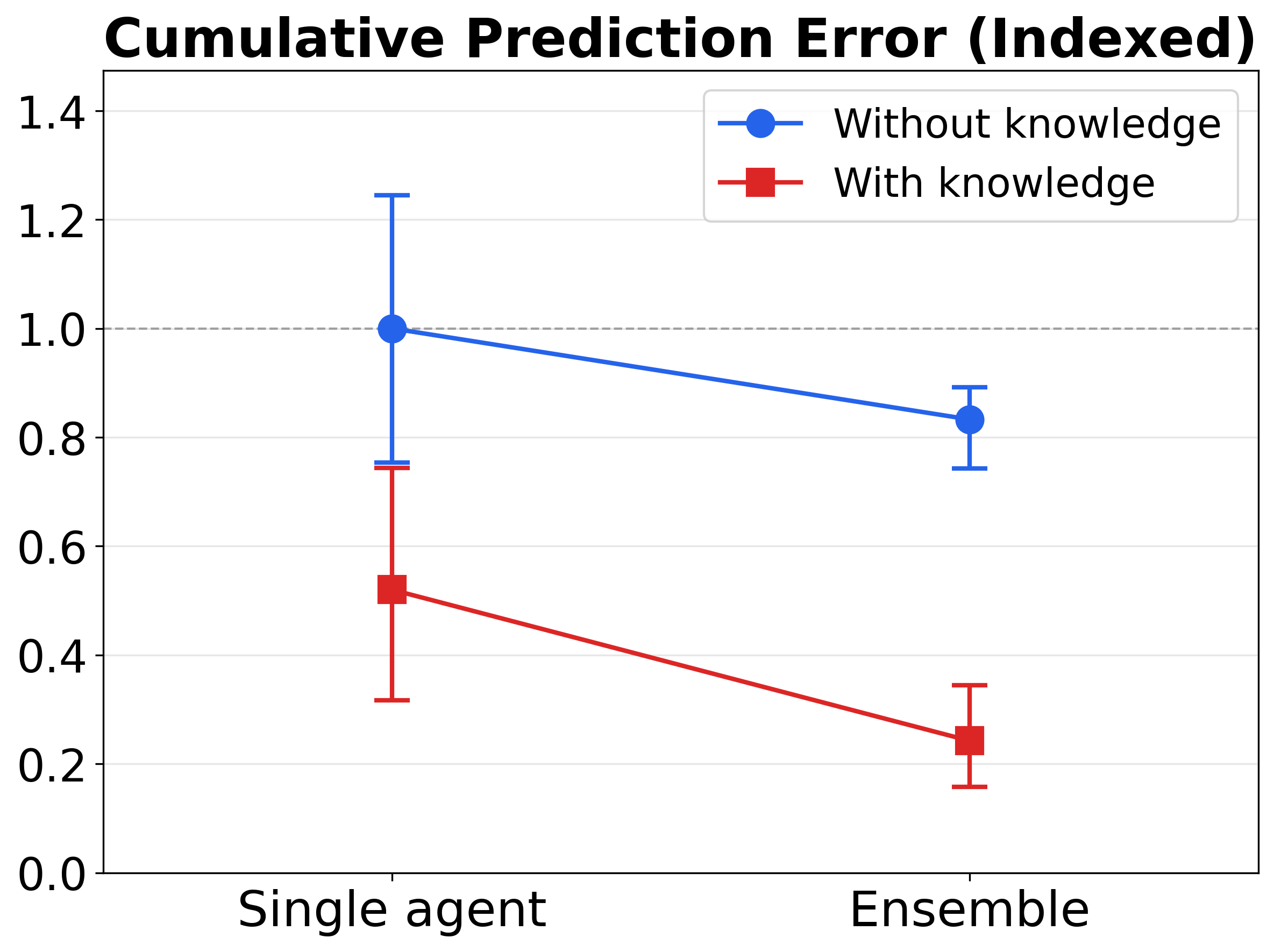}
    \caption{Cumulative prediction error}
    \label{fig:world1_error}
\end{subfigure}
\caption{
\textbf{World~1 (policy feedback only):} Performance across agent configurations. 
(a) Cumulative cases indexed to single-agent baseline; 
(b) mean reduction in transmission (lower = stronger suppression); 
(c) cumulative prediction error indexed to baseline.
}
\label{fig:world1_summary}
\end{figure*}

Aggregate results across simulation runs are summarized in Figure~\ref{fig:world1_summary}. The first panel shows that, in terms of cumulative cases, the ensemble model without knowledge performs worse than the single-agent \textit{Base} scenario, increasing total infections by roughly one third. In contrast, providing causal knowledge improves outcomes for both single-agent and ensemble settings, with the \textit{Ensemble with Knowledge} condition outperforming all others and reducing cumulative cases by approximately 50\%. The second panel shows a similar pattern for reductions in transmission, indicating that the effect of ensemble decision-making is strongly moderated by the presence of knowledge. The third panel reports cumulative absolute prediction error, which declines across conditions; the ensemble agent with knowledge produces the most accurate forecasts overall.

\subsection{World 2: Behavioral Adaptation Increases Complexity}
\label{sec:results_world2}

\begin{figure*}[t]
    \centering
    \includegraphics[width=0.8\textwidth]{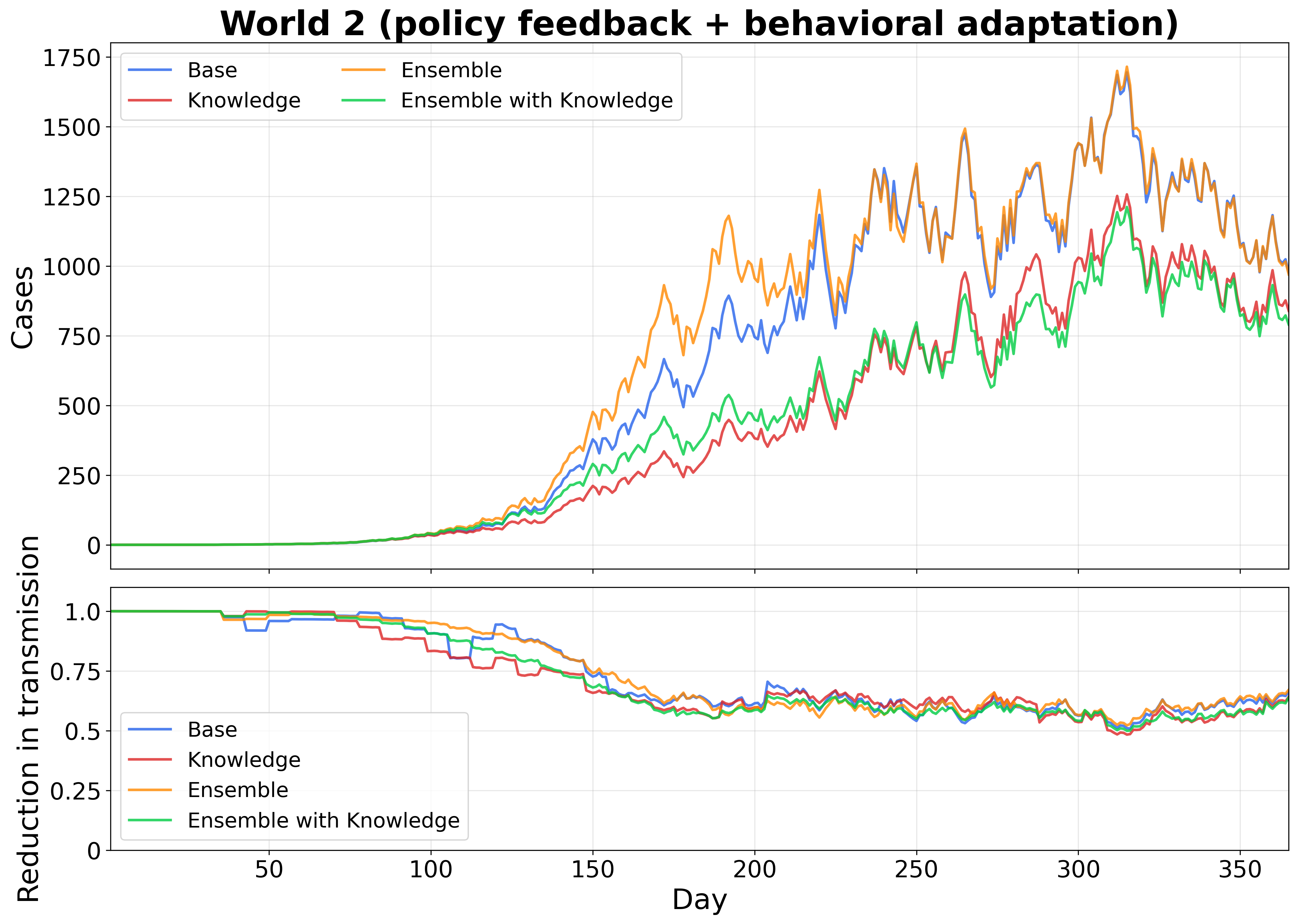}
    \caption{
    \textbf{World~2 (policy + behavioral adaptation):} Mean trajectories over 10 simulation runs.
    Top panel: reported cases; bottom panel: reduction in transmission implied by policy (lower = stronger suppression).
    }
    \label{fig:world2_timeseries}
\end{figure*}

Figure~\ref{fig:world2_timeseries} shows the corresponding World~2 results. As in World~1, all four conditions exhibit qualitatively similar reactive dynamics. However, case counts remain substantially lower across all configurations because voluntary public caution helps suppress transmission even without strong policy intervention. In simple terms, if policies are delayed, people start perceiving the threat and reacting. The context becomes more complex for the AI agent, as it cannot count on public responses remaining constant: as cases decline, the public eases their voluntary protective actions. 

In the \textit{Base} condition (blue), the AI agent exhibits largely reactive policymaking, but the outbreak is considerably dampened by behavioral adaptation. The \textit{Knowledge} condition (red) follows a similar reactive pattern but intervenes earlier and more aggressively. As a result, the transmission rate declines sooner and remains lower for a longer period over the course of the outbreak. This earlier suppression keeps case counts below the \textit{Base} trajectory, reducing year-long cumulative cases by approximately one-third.

The \textit{Ensemble without Knowledge} condition (yellow) again underperforms the \textit{Base} case. Reductions in transmission remain weak compared with other configurations, and cases slightly exceed the \textit{Base} trajectory. However, the gap is smaller than in World~1 because behavioral adaptation dampens transmission regardless of policy choices.

The \textit{Ensemble with Knowledge} configuration (green) achieves the strongest outcomes, with mean year-long cumulative cases reduced by approximately one-third relative to \textit{Base}—similar to the \textit{Knowledge} condition. Unlike in World~1, where this configuration reduced cases by half, the benefit ceiling is lower here because behavioral adaptation already provides substantial transmission suppression.

\begin{figure*}[htbp]
\centering
\begin{subfigure}[b]{0.32\textwidth}
    \includegraphics[width=\textwidth]{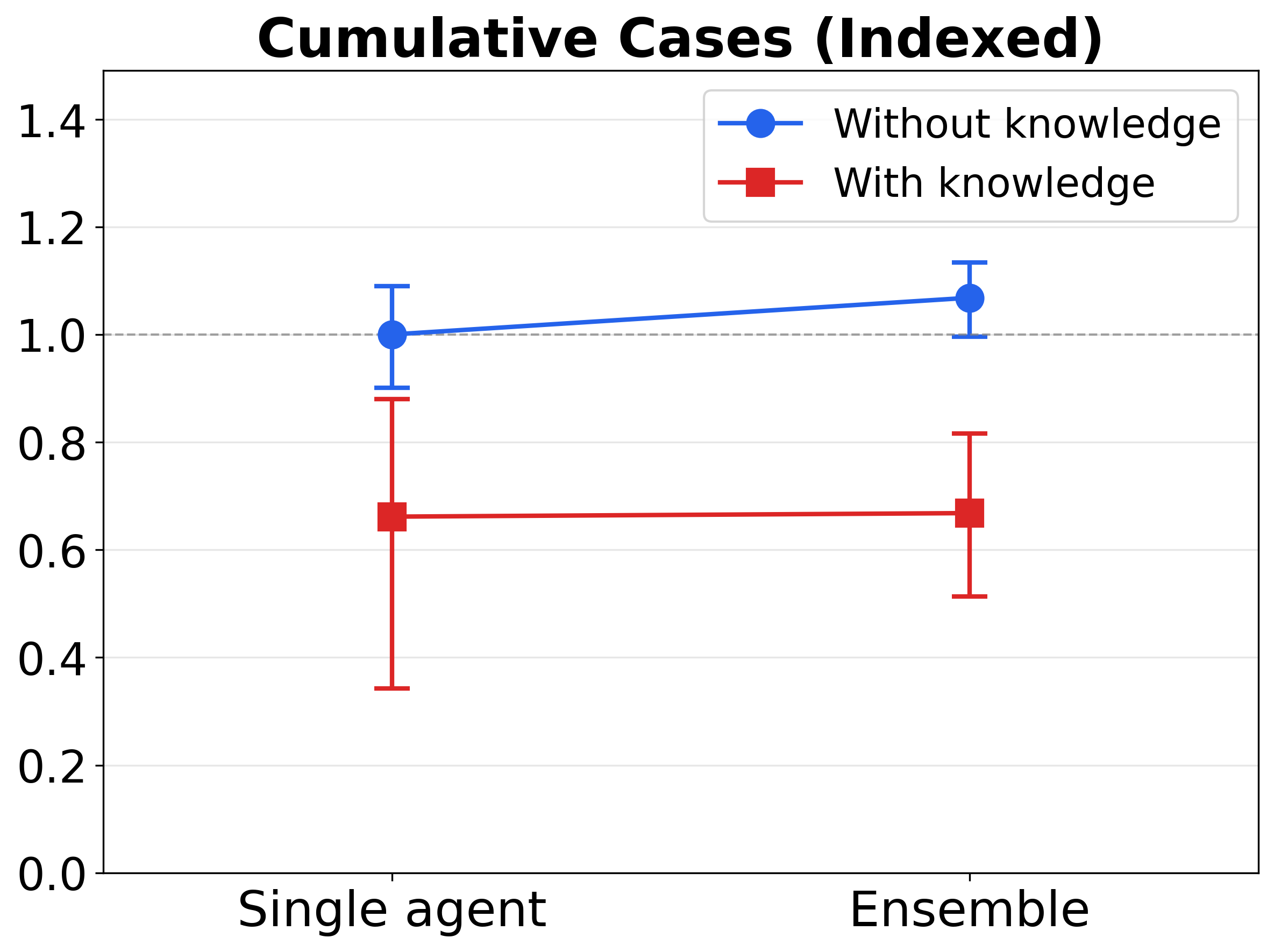}
    \caption{Cumulative cases}
    \label{fig:world2_cases}
\end{subfigure}
\hfill
\begin{subfigure}[b]{0.32\textwidth}
    \includegraphics[width=\textwidth]{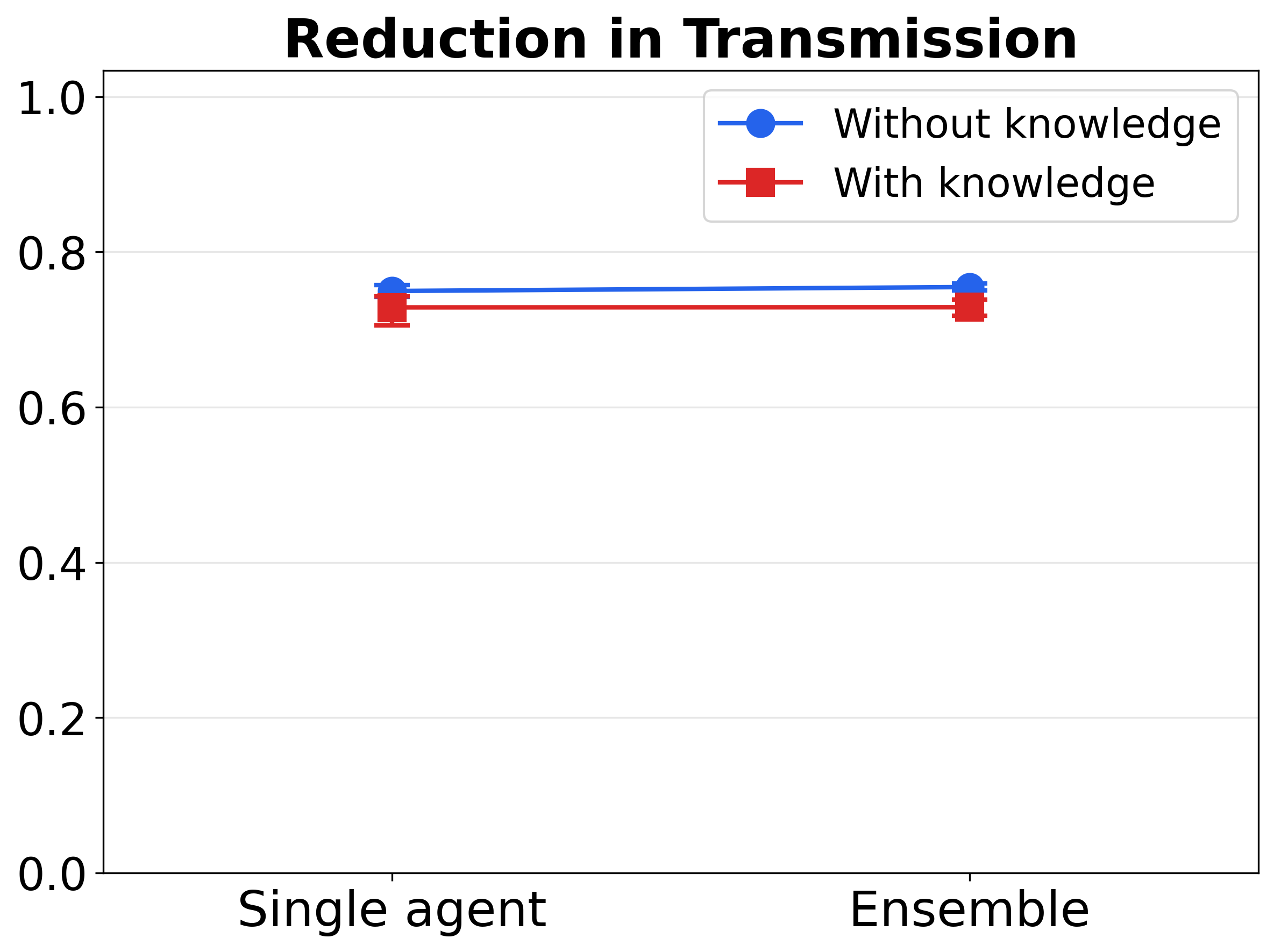}
    \caption{Reduction in transmission}
    \label{fig:world2_reduction}
\end{subfigure}
\hfill
\begin{subfigure}[b]{0.32\textwidth}
    \includegraphics[width=\textwidth]{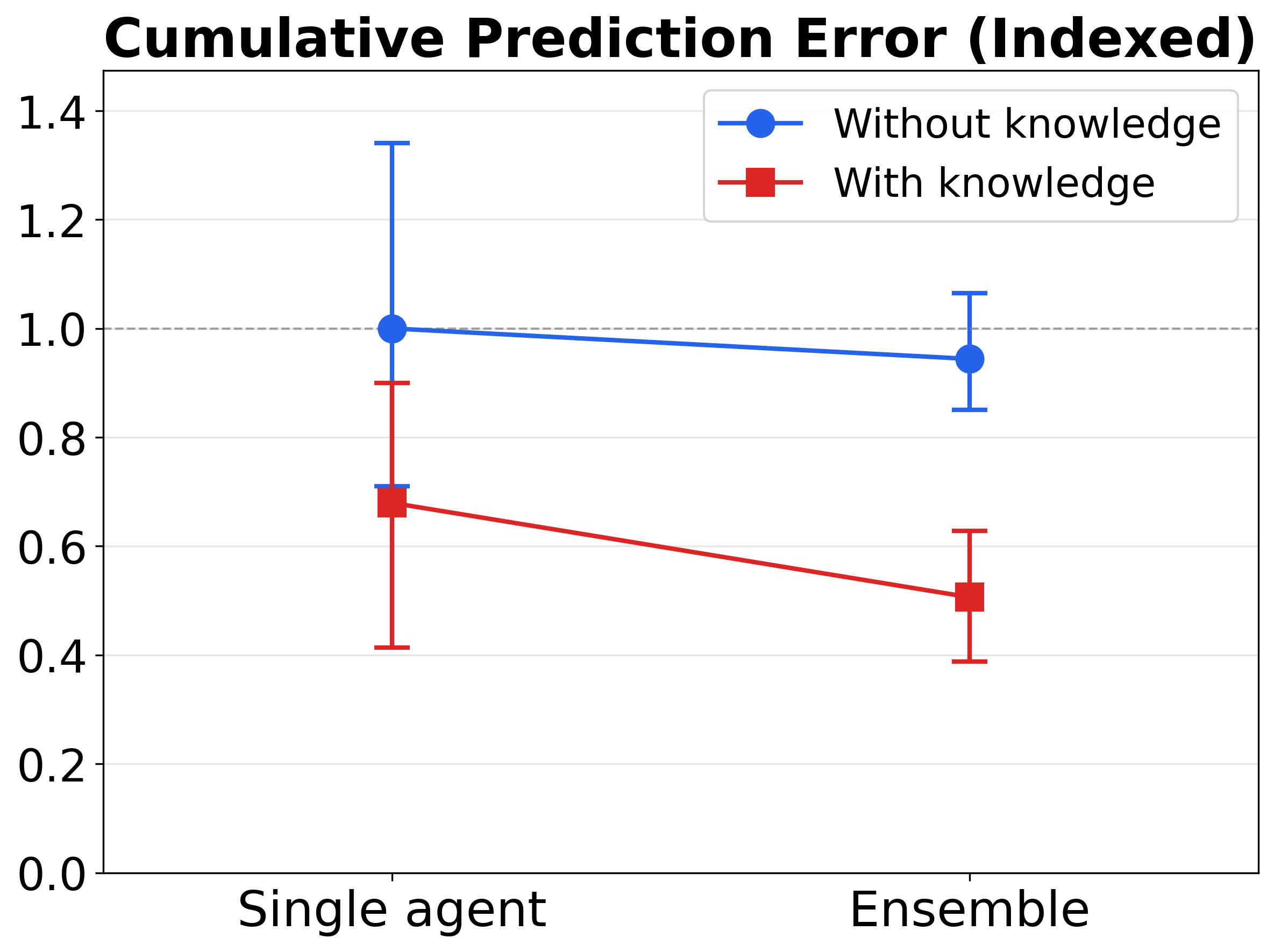}
    \caption{Cumulative prediction error}
    \label{fig:world2_error}
\end{subfigure}
\caption{
\textbf{World~2 (policy + behavioral adaptation):} Performance across agent configurations. 
(a) Cumulative cases indexed to single-agent baseline; 
(b) mean reduction in transmission (lower = stronger suppression); 
(c) cumulative prediction error indexed to baseline.
}
\label{fig:world2_summary}
\end{figure*}

Aggregate results across simulation runs are summarized in Figure~\ref{fig:world2_summary}. The first panel shows that, in terms of cumulative cases, differences between conditions are more compressed than in World~1: the \textit{Ensemble without Knowledge} condition increases total infections only modestly, while the \textit{Knowledge} and \textit{Ensemble with Knowledge} conditions both reduce cumulative cases by approximately one-third. The second panel confirms this compression: reduction in transmission values cluster tightly across all configurations, compared to the wider spread observed in World~1. The third panel reports cumulative absolute prediction error, which again declines with the Knowledge intervention; the \textit{Ensemble with Knowledge} condition produces the most accurate forecasts, followed by \textit{Knowledge} alone. This is an important observation showing that the knowledgeable agent can decline cases without necessarily imposing more restrictions but with more responsiveness leading to faster policy implementation.  

\subsection{Statistical Analysis of Agent's Decisions}
\label{sec:results_regression}

The overall trends show that agents respond to epidemic cases such that more cases lead to more restriction. In order to systematically examine this and what other factors drive the agent's predictions and decisions, we regress case predictions and business-restriction decisions on recent epidemic signals and retrieved dynamic memories. Appendix Tables~\ref{tab:reg_prediction_combined}--\ref{tab:reg_decision_combined} report the results. The results depict the agent's consistent reaction to the most recent information about the state of the epidemic. Specifically, last week's cases are strongly associated with both the agent's case predictions ($p < 0.001$) and business-restriction decisions ($p < 0.01$) across most models. 

Consistent with a belief that policy reduces transmission, in World 1, last week's business-restriction decisions enter the case prediction regression models with negative coefficients ($p < 0.001$). This relationship is weaker and less stable in World~2, where behavioral adaptation complicates attribution ($p< 0.01$).

In the business-restriction decision regression models, policy choices exhibit strong inertia: last week's decision is the strongest explanatory variable for current decision ($p < 0.001$). Recalled historical severity also shapes decisions: average cases in memory show consistently negative coefficients ($p < 0.001$), suggesting normalization or adaptation effects. In contrast, in the World~2 case prediction regressions, recalled past cases load positively, consistent with availability heuristics rather than adaptation.







\section{Discussion}

This study developed and evaluated a generative AI agent operating as a policymaker in simulated epidemic environments. The agent, powered by an LLM and equipped with recency-weighted dynamic memory, made weekly decisions about business restrictions while interacting with a mechanistic SEIR model. We examined agent performance across two environmental scenarios—one where policy was the sole behavioral lever (World~1) and another where citizens also voluntarily adapted their behavior in response to perceived risk (World~2)—and tested interventions including systems-level knowledge about feedback loops and multi-agent ensemble averaging. The Knowledge intervention consistently reduced cumulative cases by about one-third across both environments, while the Ensemble with Knowledge configuration achieved the best overall outcomes, reducing cases by half relative to the base case.

This study resonates with efforts to use AI in general, and LLMs in particular, for scientific investigation (e.g., \citep{giabbanelli2023gpt, xu_artificial_2021, Boiko2023AutonomousChemicalResearch}). Its novelty is in the introduction of the concept of AI agent as a policymaker, and showing how it depicts a human-like behavior in a simulated world, reacting to changes in risk, as well as testing ways to improve its performance. Our findings resonate with the emerging literature on generative agents and generative agent-based modeling in several ways. First, the study brings the concept of generative agents from living entities in games, such as those studied by Park et al. \citep{Park2023GenerativeAgents, Park2022SocialSimulacra} to decision-making in environments that resemble real world high-stakes policy situations. Second, whereas Williams et al. modeled citizen agents who adjust mobility in response to symptoms and prevalence information \citep{Williams2023EpidemicGABM}, we demonstrate that LLM-based agents can also operate effectively in the complementary role of policymaker—predicting trajectories, applying interventions, and adapting to feedback from their own decisions. This shift from modeling behavioral responses to modeling strategic decision-making represents a qualitative expansion of the GABM framework. Third, while studies before have applied reinforcement learning or other methods as an optimization approach in AI contexts \citep{bushaj_simulation-deep_2023}, we intentionally stay with human-like behavior paradigm of decision-making based on available information and memory. In a sense, this is consistent with human-like bounded rationality applied in the context of AI agents \citep{simon1955behavioral, simon1991bounded}. Fourth, our results speak to concerns raised about prompt sensitivity in GABM \citep{Ghaffarzadegan2024GABM, li2025let}. Rather than treating prompt sensitivity as purely a methodological challenge, we show that carefully designed knowledge interventions can systematically improve policy agent performance, suggesting that prompt engineering may serve as a tool for enhancing—not merely controlling—agent behavior.

Our work also contributes to understanding LLM limitations in strategic contexts. Our results are consistent with previous findings (e.g., \citep{Lore2024Strategic, Schmidt2024GPT35}) emphasizing the role of contextual information and framing. In our analysis, the base agent exhibits reactive rather than anticipatory policymaking, and ensemble averaging without knowledge actually worsens outcomes by amplifying policy inertia. However, our Knowledge intervention demonstrates that providing explicit causal structure can partially overcome these limitations—the agent becomes more responsive to rising cases and produces more calibrated predictions when equipped with feedback-loop explanations. This suggests that LLM-based agents may benefit from domain-specific scaffolding that compensates for their limited capacity for spontaneous causal reasoning.

These findings have implications for both research and practice. For researchers studying human decision-making in complex systems, our framework offers a controlled testbed where agent behavior can be systematically manipulated and measured—something difficult to achieve with human participants. For practitioners exploring AI-assisted decision support, our results suggest that raw LLM outputs may be insufficient for high-stakes policy contexts; effective deployment may require explicit knowledge scaffolding and careful attention to how ensemble methods interact with decision quality. Future work could extend this framework to ensemble settings where multiple policymakers interact, incorporate more realistic information delays and reporting errors, or test whether agents can learn to improve their causal models through extended experience.

Several limitations warrant consideration and provide future avenues for research. First, our SEIR(b) world model, while capturing essential epidemic dynamics, abstracts away many real-world complexities including spatial heterogeneity, age structure, healthcare capacity constraints, and economic feedbacks. Second, the agent receives perfect information about weekly case counts; real policymakers face substantial reporting delays and measurement error. Third, our Knowledge intervention provides accurate information about the true causal structure; in practice, policymakers often receive conflicting or incorrect expert advice. Fourth, we tested only one LLM (GPT-5 nano) with fixed prompting; different models or prompt formulations might yield different results. In preliminary experiments, we also evaluated GPT-4, which exhibited inconsistent responses. While GPT-5 nano provided more reliable outputs, systematic comparison across a broader range of models—including Claude and open-source alternatives—remains an important direction for future research. Fifth, our simulation does not capture many constraints that shape real-world policymaking: political feasibility, public opinion backlash, legal and constitutional limitations, coordination failures across jurisdictions, lobbying pressures, electoral cycles, and bureaucratic implementation delays. Real policymakers cannot simply choose epidemiologically optimal policies—they must navigate competing interests and institutional constraints that our agent framework does not represent. Finally, our evaluation focuses on epidemic outcomes and prediction accuracy; we do not assess whether the agent's reasoning would be perceived as legitimate or trustworthy by human stakeholders.

Despite these limitations, our study points to potentials of generative AI agents as decision makers in dynamic, feedback-rich environments. As LLMs become increasingly integrated into decision-support systems, understanding how they reason about complex systems and how their reasoning can be enhanced becomes ever more important. Our findings suggest that the path toward effective AI-assisted policymaking lies not in deploying raw model capabilities, but in thoughtfully combining AI systems with domain knowledge structures that scaffold appropriate causal reasoning. The policymaker agent developed here is a step towards continued exploration of this promising but challenging frontier.
\section{Acknowledgments}
Funding: This research is partially funded by the US National Science Foundation, Division of Mathematical Sciences and Division of Social and Economic Sciences, Award 2229819.

Data Availability: Data and Code for this study are available at \url{https://github.com/goshiaoki/AI-Agents-as-Policymakers.git}.

Competing interests: The authors have declared that they have no competing interests.

\clearpage
\appendix
\setcounter{table}{0}
\renewcommand{\thetable}{A\arabic{table}}
\setcounter{figure}{0}
\renewcommand{\thefigure}{A\arabic{figure}}

\section{Appendix}

\subsection{Simulation Algorithm}
\label{app:algorithm}

Algorithm~\ref{alg:main} presents the complete simulation procedure. The outer loop iterates over days; the inner logic handles weekly decision points. The AI-agent returns a restriction percentage in $[0, 100]$, which is converted to a fraction $G \in [0, 1]$. The \textsc{LLMDecision} function encapsulates agent configuration (knowledge intervention and ensemble averaging).

\begin{algorithm}[H]
\caption{Epidemic Simulation with AI-agent Policymaker}
\label{alg:main}
\begin{algorithmic}[1]
\REQUIRE Days $T$, decision interval $\Delta$, start week $W_s$, memory sample count $m$, scenario mode
\STATE Initialize SEIR state: $S \gets N-1$, $E \gets 0$, $I \gets 1$, $R \gets 0$
\STATE Initialize: restriction $G_0 \gets 0$, memory store $\mathcal{M} \gets []$, case buffer $\mathcal{C} \gets []$
\FOR{$t = 1$ \textbf{to} $T$}
    \STATE $w \gets \lceil t / \Delta \rceil$
    \STATE $\textit{decision\_day} \gets (t = 1)$ \textbf{or} $((t-1) \mod \Delta = 0)$
    \IF{$\textit{decision\_day}$}
        \IF{$t \leq \Delta$}
            \STATE $\bar{C}_{w-1} \gets 0$ \COMMENT{Warm-up period}
        \ELSE
            \STATE $\bar{C}_{w-1} \gets \frac{1}{\Delta} \sum_{i=t-\Delta}^{t-1} C_i$ \COMMENT{Mean of last $\Delta$ days}
        \ENDIF
        \IF{$w \geq W_s$}
            \STATE Sample memories $\mathcal{M}_m \gets \textsc{RecencyWeightedSample}(\mathcal{M}, m)$
            \STATE $(G_w, \hat{C}_0, \hat{C}_G, r) \gets \textsc{LLMDecision}(\bar{C}_{w-1}, G_{w-1}, \mathcal{M}_m)$
        \ELSE
            \STATE $G_w \gets 0$ \COMMENT{No intervention before start week}
        \ENDIF
    \ENDIF
    \STATE $g \gets 1 - \alpha G_w$
    \STATE $b \gets \textsc{BehaviorModifier}(\bar{C}_{w-1}, \text{mode})$
    \STATE $\epsilon \gets \text{Uniform}(0.5, 1.5)$
    \STATE $(S,E,I,R) \gets \textsc{SEIRStep}(S,E,I,R,\beta_0 \cdot b \cdot g \cdot \epsilon)$
    \STATE $C_t \gets E/L$; append $C_t$ to $\mathcal{C}$
    \IF{$\textit{decision\_day}$}
        \STATE Append $(w, G_w, C_t)$ to $\mathcal{M}$ \COMMENT{Store decision and realized outcome}
    \ENDIF
\ENDFOR
\end{algorithmic}
\end{algorithm}

The government effect is $g = 1 - \alpha G$. The \textsc{BehaviorModifier} function computes voluntary behavioral response based on the previous week's mean cases:
\begin{equation}
b = \begin{cases}
1 & \text{if mode} = \texttt{World 1} \\
\displaystyle \frac{1}{1 + k \bar{C}_{w-1}} & \text{if mode} = \texttt{World 2}
\end{cases}
\end{equation}
where $k$ is the behavioral sensitivity coefficient. Memories store tuples of week number, restriction level, and daily reported cases on the decision day $(w, G_w,C_t)$. Note that this daily value differs in scale from the weekly mean cases $\bar{C}_{w-1}$ provided to the agent as "officially reported cases."

\subsection{Prompt Template}
\label{app:prompt}

The prompt consists of six components assembled dynamically each week:

\begin{enumerate}
    \item \textbf{Role preamble}: Establishes the agent as a pragmatic mayor balancing public health and economic considerations.
    \item \textbf{Guiding principle}: ``The default stance should be to keep the city open. Restrictions should only be implemented when case numbers clearly justify them.''
    \item \textbf{Context block}: Current week number, last week's cases ($\bar{c}_{w-1}$), and last week's business-restriction decision level ($\pi_{w-1}$).
    \item \textbf{Memory block}: Retrieved past incidents formatted as: ``On Week $w$, the number of cases was $c$ and your shutdown level was $\pi$\%.''
    \item \textbf{Knowledge block} (optional): Scenario-specific text explaining feedback dynamics.
    \item \textbf{Output specification}: Requests JSON with four fields: \texttt{prediction\_without\_policy}, \texttt{shutdown\_percentage}, \texttt{prediction\_with\_new\_policy}, and \texttt{reasoning}.
\end{enumerate}

A baseline anchor (``During flu season, there are seven hundred flu cases every week on average'') helps the agent calibrate case magnitudes.

\subsection{Complete Prompt Example}
\label{app:full_prompt}

The following is a complete prompt as sent to the LLM API during Week 6 of a simulation run in the World 2 (behavioral adaptation) scenario with the Knowledge intervention enabled:

\begin{Verbatim}[frame=single,fontsize=\small]
You are Jennifer, the mayor of a city of one million people, facing 
an epidemic. Each week, you must decide the shutdown level for the 
city (0% to 100%), balancing public health with economic stability. 
A pragmatic, democratically elected leader in your late forties with 
a background in public policy, you focus on results over politics. 
You rely on expert advice, accurate data, and trustable models. You 
balance economic growth, public safety, transportation, and housing, 
and you explain your decisions clearly so people understand your 
reasoning. You insist on transparency and evidence-based policies, 
and you track clear metrics of success. You are ambitious yet 
cautious, knowing your choices affect real people.

Your guiding principle:
The default stance should be to keep the city open (low shutdown 
levels). Restrictions should only be implemented when case numbers 
clearly justify them.

It is early in the morning of Week 6. You must decide the shutdown 
level for Week 6.

During flu season, there are seven hundred flu cases every week on 
average.

Here are the officially reported cases:
- On Week 5, the number of cases was 1. The shutdown level was 0%.

You particularly remember the following incidents:
- On Week 1, the number of cases was 0 and your shutdown level was 0%.
- On Week 2, the number of cases was 0 and your shutdown level was 0%.
- On Week 3, the number of cases was 0 and your shutdown level was 0%.
- On Week 4, the number of cases was 0 and your shutdown level was 0%.
- On Week 5, the number of cases was 1 and your shutdown level was 0%.

Here I provide some information to help you better understand epidemic 
dynamics. Epidemics are primarily governed by interacting feedback 
loops. First, there is the reinforcing (positive) feedback loop: 
infection breeds more infection. Without intervention, cases grow 
exponentially as each infected person spreads the disease to 
susceptible individuals. The second is the balancing feedback loop 
of depletion. As more people become infected and then immune after 
recovery, the pool of susceptible individuals shrinks, which 
naturally slows transmission over time.
In addition to these biological feedback loops, there are behavioral 
feedback loops that shape transmission. As cases rise, people tend 
to grow more cautious and voluntarily adopt protective behaviors 
such as masking, distancing, and avoiding crowds. These reactions 
reduce the transmission rate. Conversely, when cases decline, 
individuals often relax their guard, which can lead to increased 
transmission and a resurgence of cases. Most importantly for your 
role, government restrictions are also part of a behavioral feedback 
loop. As you impose stricter measures on business and social 
activities, the probability of disease spread decreases. In simple 
terms, stronger shutdowns mean lower future infection rates-though 
naturally at an economic cost. Your shutdown decisions do not 
operate in isolation; they interact with voluntary citizen behavior 
driven by perceived risk.
When forecasting and making decisions, it is crucial to recognize 
that implementing or relaxing restrictive policies influences the 
spread of the disease and people's responses to those changes.

Based on the officially reported cases and your memories, you must:
Predict cases without policy: How many cases do you expect for 
Week 6 if no shutdown is implemented (0% shutdown)?
Choose your new shutdown level: What shutdown level (0-100%) should 
you implement for Week 6?
Predict cases with your new policy: How many cases do you expect for 
Week 6 after implementing your chosen shutdown level?

Output only a single JSON object with these keys:
- "prediction_without_policy": a non-negative integer representing 
  your predicted cases if no shutdown is implemented.
- "reasoning": a string with 1-2 sentences explaining your shutdown 
  decision and how changing (or maintaining) the policy affects your 
  case prediction.
- "shutdown_percentage": a number from 0 to 100 representing the new 
  shutdown level you choose. 0 is fully open. 100 is fully shut down.
- "prediction_with_new_policy": a non-negative integer representing 
  your predicted cases after implementing your new shutdown level.

Respond with JSON only:
{"prediction_without_policy": <integer>, 
 "reasoning": "<1-2 sentences explaining your logic>", 
 "shutdown_percentage": <0-100>, 
 "prediction_with_new_policy": <integer>}
\end{Verbatim}

In conditions without the Knowledge intervention, the knowledge paragraph (starting with ``Here I provide some information...'') is omitted. The memory block varies each week based on the recency-weighted retrieval described in Method section.

\subsection{Knowledge Intervention Text}
\label{app:knowledge}

The knowledge paragraph varies by scenario. In World 1 (policy feedback) scenario, the knowledge text reads as follows:

\begin{quote}
\small
\textit{``Here I provide some information to help you better understand epidemic dynamics. Epidemics are primarily governed by interacting feedback loops. First, there is the reinforcing (positive) feedback loop: infection breeds more infection. Without intervention, cases grow exponentially as each infected person spreads the disease to susceptible individuals. The second is the balancing feedback loop of depletion. As more people become infected and then immune after recovery, the pool of susceptible individuals shrinks, which naturally slows transmission over time.
Government restrictions can also affect the spread of the disease by influencing people's behavior. As you impose stricter measures on business and social activities, the probability of disease spread decreases. In simple terms, stronger shutdowns mean lower future infection rates—though naturally at an economic cost. In your town people ignore the disease unless the government imposes restrictions which they will comply with.
When forecasting and making decisions, it is crucial to recognize that implementing or relaxing restrictive policies influences the spread of the disease.''}
\end{quote}

\subsection{Computational Notes}

Simulations were conducted using OpenAI's API (gpt-5-nano, accessed December 15, 2025). The GPT-5 nano model does not support a configurable temperature parameter; all API calls use a fixed temperature of 1.0, which contributes to output stochasticity. Each 365-day simulation with Ensemble requires approximately 520 API calls (52 decision weeks $\times$ 10 inferences). The 10-run batch for each of the 8 conditions (2 worlds $\times$ 4 configurations) totals 80 simulation runs.

Random seeds (set to 42) ensure reproducibility of epidemic dynamics and memory sampling. LLM outputs remain stochastic due to API-side sampling, contributing to across-run variability even with identical seeds.

\subsection{Multi-inference averaging}
\label{app:averaging}

In the \textit{Ensemble} conditions, the agent makes 10 independent LLM calls on each decision week and averages the numeric fields:
\texttt{shutdown\_percentage}, \texttt{prediction\_without\_policy}, and \texttt{prediction\_with\_new\_policy}. The simulation records one representative \texttt{reasoning} string (from a sampled response) to preserve interpretability.

\subsection{Example LLM Output}
\label{app:output}

The following is an example JSON response from the LLM corresponding to the prompt:

\begin{Verbatim}[frame=single,fontsize=\small]
{"prediction_without_policy": 3, 
 "reasoning": "I am maintaining a 0% shutdown for Week 6 because last 
 week had only 1 reported case and there is no clear surge; keeping 
 the city open minimizes economic disruption. With zero restrictions, 
 I forecast about 3 cases in Week 6; imposing even a small shutdown 
 would reduce transmission and lower Week 6 cases, but at an 
 unnecessary economic cost given the current data.", 
 "shutdown_percentage": 0, 
 "prediction_with_new_policy": 3}
\end{Verbatim}

In this example, the agent predicts 3 cases regardless of policy, chooses to maintain a 0\% shutdown level, and provides reasoning that reflects the guiding principle of keeping the city open when case numbers are low. In the Ensemble conditions, the numeric fields (\texttt{prediction\_without\_policy}, \texttt{shutdown\_percentage}, and \texttt{prediction\_with\_new\_policy}) are averaged across 10 independent API calls, while one representative \texttt{reasoning} string is retained.

\begin{table}[htbp]
\centering
\small
\caption{Combined regression analysis of case predictions. All conditions pooled within each environment. Model~1 includes only recent information (last week's cases and business-restriction decision). Model~2 adds dynamic memory variables (average business-restriction decision and cases stored in memory). Model~3 further adds treatment effects (Knowledge, ensemble, and their interaction).}
\label{tab:reg_prediction_combined}
\begin{tabular}{lcccccc}
\toprule
 & \multicolumn{3}{c}{World 1 (Policy Feedback)} & \multicolumn{3}{c}{World 2 (Behavioral Adaptation)} \\
\cmidrule(lr){2-4} \cmidrule(lr){5-7}
 & Model 1 & Model 2 & Model 3 & Model 1 & Model 2 & Model 3 \\
\midrule
\textit{Recent Information} & & & & & & \\[0.3em]
Last Week Cases & 0.972*** & 1.027*** & 1.013*** & 0.842*** & 0.721*** & 0.690*** \\
 & (0.005) & (0.019) & (0.019) & (0.008) & (0.037) & (0.036) \\[0.5em]
Last Week Decision & $-$292.2*** & $-$481.6*** & $-$450.5*** & $-$244.7*** & $-$166.2** & $-$145.6** \\
 & (37.7) & (73.7) & (72.7) & (26.9) & (48.5) & (47.9) \\[0.8em]
\textit{Dynamic Memory} & & & & & & \\[0.3em]
Avg Decision in Memory & & 235.4** & 409.5*** & & $-$100.4 & 27.4 \\
 & & (83.6) & (85.6) & & (55.8) & (57.4) \\[0.5em]
Avg Cases in Memory & & $-$0.061** & $-$0.069*** & & 0.126** & 0.140*** \\
 & & (0.019) & (0.019) & & (0.037) & (0.036) \\[0.8em]
\textit{Treatment Effects} & & & & & & \\[0.3em]
Knowledge & & & $-$46.6* & & & $-$61.3*** \\
 & & & (22.0) & & & (11.6) \\[0.5em]
Ensemble & & & 105.7*** & & & 13.2 \\
 & & & (22.2) & & & (10.7) \\[0.5em]
Ensemble $\times$ Knowledge & & & $-$148.0*** & & & $-$18.4 \\
 & & & (31.4) & & & (15.1) \\[0.5em]
Constant & 218.6*** & 213.7*** & 202.2*** & 225.1*** & 229.6*** & 248.2*** \\
 & (12.9) & (13.2) & (18.4) & (7.1) & (7.3) & (9.5) \\
\midrule
$N$ & 1920 & 1920 & 1920 & 1920 & 1920 & 1920 \\
$R^2$ & 0.948 & 0.949 & 0.950 & 0.860 & 0.861 & 0.865 \\
Adj.\ $R^2$ & 0.948 & 0.949 & 0.950 & 0.860 & 0.861 & 0.865 \\
\bottomrule
\multicolumn{7}{l}{\footnotesize \textit{Note}: Standard errors in parentheses. ***$p<0.001$, **$p<0.01$, *$p<0.05$.} \\
\end{tabular}
\end{table}


\begin{table}[htbp]
\centering
\small
\caption{Combined regression analysis of business-restriction decisions. All conditions pooled within each environment. Model~1 includes only recent information (last week's cases and business-restriction decision). Model~2 adds dynamic memory variables (average business-restriction decisions and cases stored in memory). Model~3 further adds treatment effects (Knowledge, Ensemble, and their interaction).}
\label{tab:reg_decision_combined}
\begin{tabular}{lcccccc}
\toprule
 & \multicolumn{3}{c}{World 1 (Policy Feedback)} & \multicolumn{3}{c}{World 2 (Behavioral Adaptation)} \\
\cmidrule(lr){2-4} \cmidrule(lr){5-7}
 & Model 1 & Model 2 & Model 3 & Model 1 & Model 2 & Model 3 \\
\midrule
\textit{Recent Information} & & & & & & \\[0.3em]
Last Week Cases ($\times 10^{-5}$) & 0.17 & 6.90*** & 7.31*** & $-$1.28** & 13.56*** & 15.07*** \\
 & (0.19) & (0.65) & (0.65) & (0.44) & (1.95) & (1.94) \\[0.5em]
Last Week Decision & 0.791*** & 0.493*** & 0.485*** & 0.745*** & 0.497*** & 0.487*** \\
 & (0.014) & (0.026) & (0.025) & (0.015) & (0.026) & (0.026) \\[0.8em]
\textit{Dynamic Memory} & & & & & & \\[0.3em]
Avg Decision in Memory & & 0.376*** & 0.327*** & & 0.333*** & 0.271*** \\
 & & (0.029) & (0.030) & & (0.030) & (0.031) \\[0.5em]
Avg Cases in Memory ($\times 10^{-5}$) & & $-$7.60*** & $-$7.40*** & & $-$15.86*** & $-$16.52*** \\
 & & (0.66) & (0.66) & & (1.97) & (1.95) \\[0.8em]
\textit{Treatment Effects} & & & & & & \\[0.3em]
Knowledge & & & 0.019* & & & 0.031*** \\
 & & & (0.008) & & & (0.006) \\[0.5em]
Ensemble & & & $-$0.021** & & & $-$0.006 \\
 & & & (0.008) & & & (0.006) \\[0.5em]
Ensemble $\times$ Knowledge & & & 0.035** & & & 0.007 \\
 & & & (0.011) & & & (0.008) \\[0.5em]
Constant & 0.051*** & 0.043*** & 0.040*** & 0.045*** & 0.034*** & 0.024*** \\
 & (0.005) & (0.005) & (0.006) & (0.004) & (0.004) & (0.005) \\
\midrule
$N$ & 1920 & 1920 & 1920 & 1920 & 1920 & 1920 \\
$R^2$ & 0.632 & 0.680 & 0.686 & 0.567 & 0.606 & 0.616 \\
Adj.\ $R^2$ & 0.632 & 0.679 & 0.685 & 0.566 & 0.605 & 0.614 \\
\bottomrule
\multicolumn{7}{l}{\footnotesize \textit{Note}: Standard errors in parentheses. ***$p<0.001$, **$p<0.01$, *$p<0.05$.} \\
\end{tabular}
\end{table}

\clearpage

\bibliographystyle{unsrtnat}
\bibliography{references}

\begin{thebibliography}{40}
\providecommand{\natexlab}[1]{#1}
\providecommand{\url}[1]{\texttt{#1}}
\expandafter\ifx\csname urlstyle\endcsname\relax
  \providecommand{\doi}[1]{doi: #1}\else
  \providecommand{\doi}{doi: \begingroup \urlstyle{rm}\Url}\fi

\bibitem[Wooldridge and Jennings(1995)]{Wooldridge_Jennings_1995}
Michael Wooldridge and Nicholas~R. Jennings.
\newblock Intelligent agents: theory and practice.
\newblock \emph{The Knowledge Engineering Review}, 10\penalty0 (2):\penalty0
  115–152, 1995.
\newblock \doi{10.1017/S0269888900008122}.

\bibitem[Park et~al.(2023)Park, O'Brien, Cai, Morris, Liang, and
  Bernstein]{Park2023GenerativeAgents}
Joon~Sung Park, Joseph~C. O'Brien, Carrie~J. Cai, Meredith~Ringel Morris, Percy
  Liang, and Michael~S. Bernstein.
\newblock Generative agents: Interactive simulacra of human behavior.
\newblock In \emph{Proceedings of the 36th Annual ACM Symposium on User
  Interface Software and Technology (UIST '23)}, pages 2:1--2:22, New York, NY,
  USA, 2023. Association for Computing Machinery.
\newblock \doi{10.1145/3586183.3606763}.
\newblock URL \url{https://doi.org/10.1145/3586183.3606763}.

\bibitem[Chong et~al.(2021)Chong, Yu, Keeling, and
  Ruyter]{chong_ai-chatbots_2021}
Terrence Chong, Ting Yu, Debbie~Isobel Keeling, and Ko~de Ruyter.
\newblock {AI}-chatbots on the services frontline addressing the challenges and
  opportunities of agency.
\newblock \emph{Journal of Retailing and Consumer Services}, 63:\penalty0
  102735, 2021.
\newblock ISSN 0969-6989.
\newblock \doi{https://doi.org/10.1016/j.jretconser.2021.102735}.
\newblock URL
  \url{https://www.sciencedirect.com/science/article/pii/S0969698921003015}.

\bibitem[Knani et~al.(2022)Knani, Echchakoui, and
  Ladhari]{knani_artificial_2022}
Mouna Knani, Said Echchakoui, and Riadh Ladhari.
\newblock Artificial intelligence in tourism and hospitality: {Bibliometric}
  analysis and research agenda.
\newblock \emph{International Journal of Hospitality Management}, 107:\penalty0
  103317, 2022.
\newblock ISSN 0278-4319.
\newblock \doi{https://doi.org/10.1016/j.ijhm.2022.103317}.
\newblock URL
  \url{https://www.sciencedirect.com/science/article/pii/S0278431922001797}.

\bibitem[Xu et~al.(2024)Xu, Sun, Ren, Guo, Pan, Lin, Sun, and Han]{xu_ai_2024}
Ruoxi Xu, Yingfei Sun, Mengjie Ren, Shiguang Guo, Ruotong Pan, Hongyu Lin,
  Le~Sun, and Xianpei Han.
\newblock {AI} for social science and social science of {AI}: {A} survey.
\newblock \emph{Information Processing \& Management}, 61\penalty0
  (3):\penalty0 103665, 2024.
\newblock ISSN 0306-4573.
\newblock \doi{https://doi.org/10.1016/j.ipm.2024.103665}.
\newblock URL
  \url{https://www.sciencedirect.com/science/article/pii/S0306457324000256}.

\bibitem[Xi et~al.(2025)Xi, Chen, Guo, He, Ding, Hong, Zhang, Wang, Jin, Zhou,
  Zheng, Fan, Wang, Xiong, Zhou, Wang, Jiang, Zou, Liu, Yin, Dou, Weng, Qin,
  Zheng, Qiu, Huang, Zhang, and Gui]{xi_rise_2025}
Zhiheng Xi, Wenxiang Chen, Xin Guo, Wei He, Yiwen Ding, Boyang Hong, Ming
  Zhang, Junzhe Wang, Senjie Jin, Enyu Zhou, Rui Zheng, Xiaoran Fan, Xiao Wang,
  Limao Xiong, Yuhao Zhou, Weiran Wang, Changhao Jiang, Yicheng Zou, Xiangyang
  Liu, Zhangyue Yin, Shihan Dou, Rongxiang Weng, Wenjuan Qin, Yongyan Zheng,
  Xipeng Qiu, Xuanjing Huang, Qi~Zhang, and Tao Gui.
\newblock The rise and potential of large language model based agents: a
  survey.
\newblock \emph{Science China Information Sciences}, 68\penalty0 (2):\penalty0
  121101, January 2025.
\newblock ISSN 1869-1919.
\newblock \doi{10.1007/s11432-024-4222-0}.
\newblock URL \url{https://doi.org/10.1007/s11432-024-4222-0}.

\bibitem[Wang et~al.(2024)Wang, Ma, Feng, Zhang, Yang, Zhang, Chen, Tang, Chen,
  Lin, Zhao, Wei, and Wen]{wang_survey_2024}
Lei Wang, Chen Ma, Xueyang Feng, Zeyu Zhang, Hao Yang, Jingsen Zhang, Zhiyuan
  Chen, Jiakai Tang, Xu~Chen, Yankai Lin, Wayne~Xin Zhao, Zhewei Wei, and
  Jirong Wen.
\newblock A survey on large language model based autonomous agents.
\newblock \emph{Frontiers of Computer Science}, 18\penalty0 (6):\penalty0
  186345, March 2024.
\newblock ISSN 2095-2236.
\newblock \doi{10.1007/s11704-024-40231-1}.
\newblock URL \url{https://doi.org/10.1007/s11704-024-40231-1}.

\bibitem[Gao et~al.(2024)Gao, Lan, Li, Yuan, Ding, Zhou, Xu, and
  Li]{gao_large_2024}
Chen Gao, Xiaochong Lan, Nian Li, Yuan Yuan, Jingtao Ding, Zhilun Zhou, Fengli
  Xu, and Yong Li.
\newblock Large language models empowered agent-based modeling and simulation:
  a survey and perspectives.
\newblock \emph{Humanities and Social Sciences Communications}, 11\penalty0
  (1):\penalty0 1259, September 2024.
\newblock ISSN 2662-9992.
\newblock \doi{10.1057/s41599-024-03611-3}.
\newblock URL \url{https://doi.org/10.1057/s41599-024-03611-3}.

\bibitem[Bail(2024)]{bail2024can}
Christopher~A. Bail.
\newblock Can generative ai improve social science?
\newblock \emph{Proceedings of the National Academy of Sciences}, 121\penalty0
  (21):\penalty0 e2314021121, 2024.
\newblock \doi{10.1073/pnas.2314021121}.
\newblock URL \url{https://www.pnas.org/doi/abs/10.1073/pnas.2314021121}.

\bibitem[Ziems et~al.(2024)Ziems, Held, Shaikh, Chen, Zhang, and
  Yang]{ziems_can_2024}
Caleb Ziems, William Held, Omar Shaikh, Jiaao Chen, Zhehao Zhang, and Diyi
  Yang.
\newblock Can {Large} {Language} {Models} {Transform} {Computational} {Social}
  {Science}?
\newblock \emph{Computational Linguistics}, 50\penalty0 (1):\penalty0 237--291,
  March 2024.
\newblock ISSN 0891-2017.
\newblock \doi{10.1162/coli_a_00502}.
\newblock URL \url{https://doi.org/10.1162/coli_a_00502}.
\newblock \_eprint:
  https://direct.mit.edu/coli/article-pdf/50/1/237/2367175/coli\_a\_00502.pdf.

\bibitem[Park et~al.(2022)Park, Popowski, Cai, Morris, Liang, and
  Bernstein]{Park2022SocialSimulacra}
Joon~Sung Park, Lindsay Popowski, Carrie~J. Cai, Meredith~Ringel Morris, Percy
  Liang, and Michael~S. Bernstein.
\newblock Social simulacra: Creating populated prototypes for social computing
  systems.
\newblock In \emph{Proceedings of the 35th Annual ACM Symposium on User
  Interface Software and Technology (UIST '22)}, pages 74:1--74:18, New York,
  NY, USA, 2022. Association for Computing Machinery.
\newblock \doi{10.1145/3526113.3545616}.
\newblock URL \url{https://doi.org/10.1145/3526113.3545616}.

\bibitem[Park et~al.(2024)Park, Zou, Shaw, Hill, Cai, Morris, Willer, Liang,
  and Bernstein]{Park2024ThousandPeople}
Joon~Sung Park, Carolyn~Q. Zou, Aaron Shaw, Benjamin~Mako Hill, Carrie~J. Cai,
  Meredith~Ringel Morris, Robb Willer, Percy Liang, and Michael~S. Bernstein.
\newblock Generative agent simulations of 1,000 people.
\newblock \emph{arXiv}, 2024.
\newblock \doi{10.48550/arXiv.2411.10109}.
\newblock URL \url{https://arxiv.org/abs/2411.10109}.

\bibitem[Argyle et~al.(2023)Argyle, Busby, Fulda, Gubler, Rytting, and
  Wingate]{Argyle2023OutOfOneMany}
Lisa~P. Argyle, Ethan~C. Busby, Nancy Fulda, Joshua~R. Gubler, Cameron Rytting,
  and David Wingate.
\newblock Out of one, many: Using language models to simulate human samples.
\newblock \emph{Political Analysis}, 31\penalty0 (3):\penalty0 337--351, 2023.
\newblock \doi{10.1017/pan.2023.2}.
\newblock URL \url{https://doi.org/10.1017/pan.2023.2}.

\bibitem[Giabbanelli et~al.(2025)Giabbanelli, Daumas, Flandre, Pitkar, and
  Vazquez-Estrada]{Giabbanelli03092025}
Philippe~J. Giabbanelli, Cedric Daumas, Noe~Y. Flandre, Ashutosh Pitkar, and
  Jessica Vazquez-Estrada.
\newblock Promoting empathy in decision-making by turning agent-based models
  into stories using large-language models.
\newblock \emph{Journal of Simulation}, 0\penalty0 (0):\penalty0 1--21, 2025.
\newblock \doi{10.1080/17477778.2025.2536663}.
\newblock URL \url{https://doi.org/10.1080/17477778.2025.2536663}.

\bibitem[Ghaffarzadegan et~al.(2024)Ghaffarzadegan, Majumdar, Williams, and
  Hosseinichimeh]{Ghaffarzadegan2024GABM}
Navid Ghaffarzadegan, Arindam Majumdar, Ross Williams, and Niyousha
  Hosseinichimeh.
\newblock Generative agent-based modeling: an introduction and tutorial.
\newblock \emph{System Dynamics Review}, 40\penalty0 (4):\penalty0 e1761, 2024.
\newblock \doi{10.1002/sdr.1761}.
\newblock URL \url{https://doi.org/10.1002/sdr.1761}.

\bibitem[Williams et~al.(2023)Williams, Hosseinichimeh, Majumdar, and
  Ghaffarzadegan]{Williams2023EpidemicGABM}
Ross Williams, Niyousha Hosseinichimeh, Arindam Majumdar, and Navid
  Ghaffarzadegan.
\newblock Epidemic modeling with generative agents.
\newblock \emph{arXiv}, 2023.
\newblock \doi{10.48550/arXiv.2307.04986}.
\newblock URL \url{https://arxiv.org/abs/2307.04986}.

\bibitem[Lor\`{e} and Heydari(2024)]{Lore2024Strategic}
Nunzio Lor\`{e} and Babak Heydari.
\newblock Strategic behavior of large language models and the role of game
  structure versus contextual framing.
\newblock \emph{Scientific Reports}, 14\penalty0 (1):\penalty0 18490, August
  2024.
\newblock ISSN 2045-2322.
\newblock \doi{10.1038/s41598-024-69032-z}.
\newblock URL \url{https://www.nature.com/articles/s41598-024-69032-z}.
\newblock Publisher: Nature Publishing Group.

\bibitem[Schmidt et~al.(2024)Schmidt, Bonati, Köbis, and
  Soraperra]{Schmidt2024GPT35}
Eva-Madeleine Schmidt, Sara Bonati, Nils Köbis, and Ivan Soraperra.
\newblock {GPT}-3.5 altruistic advice is sensitive to reciprocal concerns but
  not to strategic risk.
\newblock \emph{Scientific Reports}, 14\penalty0 (1):\penalty0 22274, September
  2024.
\newblock ISSN 2045-2322.
\newblock \doi{10.1038/s41598-024-73306-x}.
\newblock URL \url{https://doi.org/10.1038/s41598-024-73306-x}.

\bibitem[Akata et~al.(2025)Akata, Schulz, Coda-Forno, Oh, Bethge, and
  Schulz]{Akata2023RepeatedGames}
Elif Akata, Lion Schulz, Julian Coda-Forno, Seong~Joon Oh, Matthias Bethge, and
  Eric Schulz.
\newblock Playing repeated games with large language models.
\newblock \emph{Nature Human Behaviour}, 9\penalty0 (7):\penalty0 1380--1390,
  July 2025.
\newblock ISSN 2397-3374.
\newblock \doi{10.1038/s41562-025-02172-y}.
\newblock URL \url{https://doi.org/10.1038/s41562-025-02172-y}.

\bibitem[Gao et~al.(2023)Gao, Lan, Lu, Mao, Piao, Wang, Jin, and
  Li]{Gao2023SocialNetwork}
Chen Gao, Xiaochong Lan, Zhihong Lu, Jinzhu Mao, Jinghua Piao, Huandong Wang,
  Depeng Jin, and Yong Li.
\newblock S3: Social-network simulation system with large language
  model-empowered agents.
\newblock \emph{arXiv}, 2023.
\newblock \doi{10.48550/arXiv.2307.14984}.
\newblock URL \url{https://arxiv.org/abs/2307.14984}.

\bibitem[Xu et~al.(2020)Xu, Shieh, Esch, and Ling]{xu_ai_2020}
Yingzi Xu, Chih-Hui Shieh, Patrick~van Esch, and I.-Ling Ling.
\newblock {AI} customer service: {Task} complexity, problem-solving ability,
  and usage intention.
\newblock \emph{Australasian Marketing Journal (AMJ)}, 28\penalty0
  (4):\penalty0 189--199, 2020.
\newblock ISSN 1441-3582.
\newblock \doi{https://doi.org/10.1016/j.ausmj.2020.03.005}.
\newblock URL
  \url{https://www.sciencedirect.com/science/article/pii/S1441358220300240}.

\bibitem[(FAIR)† et~al.(2022)(FAIR)†, Bakhtin, Brown, Dinan, Farina,
  Flaherty, Fried, Goff, Gray, Hu, Jacob, Komeili, Konath, Kwon, Lerer, Lewis,
  Miller, Mitts, Renduchintala, Roller, Rowe, Shi, Spisak, Wei, Wu, Zhang, and
  Zijlstra]{meta2022human}
Meta Fundamental AI Research Diplomacy~Team (FAIR)†, Anton Bakhtin, Noam
  Brown, Emily Dinan, Gabriele Farina, Colin Flaherty, Daniel Fried, Andrew
  Goff, Jonathan Gray, Hengyuan Hu, Athul~Paul Jacob, Mojtaba Komeili, Karthik
  Konath, Minae Kwon, Adam Lerer, Mike Lewis, Alexander~H. Miller, Sasha Mitts,
  Adithya Renduchintala, Stephen Roller, Dirk Rowe, Weiyan Shi, Joe Spisak,
  Alexander Wei, David Wu, Hugh Zhang, and Markus Zijlstra.
\newblock Human-level play in the game of <i>diplomacy</i> by combining
  language models with strategic reasoning.
\newblock \emph{Science}, 378\penalty0 (6624):\penalty0 1067--1074, 2022.
\newblock \doi{10.1126/science.ade9097}.
\newblock URL \url{https://www.science.org/doi/abs/10.1126/science.ade9097}.

\bibitem[Espinoza et~al.(2024)Espinoza, Saad-Roy, Grenfell, Levin, and
  Marathe]{espinoza_adaptive_2024}
Baltazar Espinoza, Chadi~M. Saad-Roy, Bryan~T. Grenfell, Simon~A. Levin, and
  Madhav Marathe.
\newblock Adaptive human behaviour modulates the impact of immune life history
  and vaccination on long-term epidemic dynamics.
\newblock \emph{Proceedings of the Royal Society B: Biological Sciences},
  291\penalty0 (2033):\penalty0 20241772, October 2024.
\newblock ISSN 0962-8452.
\newblock \doi{10.1098/rspb.2024.1772}.
\newblock URL \url{https://doi.org/10.1098/rspb.2024.1772}.

\bibitem[Pant et~al.(2024)Pant, Safdar, Santillana, and
  Gumel]{pant_mathematical_2024}
Binod Pant, Salman Safdar, Mauricio Santillana, and Abba~B. Gumel.
\newblock Mathematical {Assessment} of the {Role} of {Human} {Behavior}
  {Changes} on {SARS}-{CoV}-2 {Transmission} {Dynamics} in the {United}
  {States}.
\newblock \emph{Bulletin of Mathematical Biology}, 86\penalty0 (8):\penalty0
  92, June 2024.
\newblock ISSN 1522-9602.
\newblock \doi{10.1007/s11538-024-01324-x}.
\newblock URL \url{https://doi.org/10.1007/s11538-024-01324-x}.

\bibitem[Fenichel et~al.(2011)Fenichel, Castillo-Chavez, Ceddia, Chowell,
  Parra, Hickling, Holloway, Horan, Morin, Perrings, Springborn, Velazquez, and
  Villalobos]{fenichel_adaptive_2011}
Eli~P. Fenichel, Carlos Castillo-Chavez, M.~G. Ceddia, Gerardo Chowell, Paula
  A.~Gonzalez Parra, Graham~J. Hickling, Garth Holloway, Richard Horan,
  Benjamin Morin, Charles Perrings, Michael Springborn, Leticia Velazquez, and
  Cristina Villalobos.
\newblock Adaptive human behavior in epidemiological models.
\newblock \emph{Proceedings of the National Academy of Sciences}, 108\penalty0
  (15):\penalty0 6306--6311, 2011.
\newblock \doi{10.1073/pnas.1011250108}.
\newblock URL \url{https://www.pnas.org/doi/abs/10.1073/pnas.1011250108}.
\newblock \_eprint: https://www.pnas.org/doi/pdf/10.1073/pnas.1011250108.

\bibitem[Kermack and McKendrick(1927)]{kermack_contribution_1927}
William~Ogilvy Kermack and A.~G. McKendrick.
\newblock A contribution to the mathematical theory of epidemics.
\newblock \emph{Proceedings of the Royal Society of London. Series A,
  Containing Papers of a Mathematical and Physical Character}, 115\penalty0
  (772):\penalty0 700--721, August 1927.
\newblock ISSN 0950-1207.
\newblock \doi{10.1098/rspa.1927.0118}.
\newblock URL \url{https://doi.org/10.1098/rspa.1927.0118}.

\bibitem[Ross and Hudson(1917)]{ross_application_1917}
Ronald Ross and Hilda~P. Hudson.
\newblock An application of the theory of probabilities to the study of a
  priori pathometry.—{Part} {II}.
\newblock \emph{Proceedings of the Royal Society of London. Series A,
  Containing Papers of a Mathematical and Physical Character}, 93\penalty0
  (650):\penalty0 212--225, May 1917.
\newblock ISSN 0950-1207.
\newblock \doi{10.1098/rspa.1917.0014}.
\newblock URL \url{https://doi.org/10.1098/rspa.1917.0014}.

\bibitem[Ferguson(2007)]{ferguson_capturing_2007}
Neil Ferguson.
\newblock Capturing human behaviour.
\newblock \emph{Nature}, 446\penalty0 (7137):\penalty0 733--733, April 2007.
\newblock ISSN 1476-4687.
\newblock \doi{10.1038/446733a}.
\newblock URL \url{https://doi.org/10.1038/446733a}.

\bibitem[{OpenAI}(2025)]{OpenAI2025GPT5}
{OpenAI}.
\newblock Introducing {GPT-5} for developers.
\newblock Technical report, OpenAI, August 2025.
\newblock URL \url{https://openai.com/index/introducing-gpt-5-for-developers/}.

\bibitem[Rahmandad et~al.(2022)Rahmandad, Xu, and
  Ghaffarzadegan]{rahmandad_enhancing_2022}
Hazhir Rahmandad, Ran Xu, and Navid Ghaffarzadegan.
\newblock Enhancing long-term forecasting: {Learning} from {COVID}-19 models.
\newblock \emph{PLOS Computational Biology}, 18\penalty0 (5):\penalty0 1--15,
  May 2022.
\newblock \doi{10.1371/journal.pcbi.1010100}.
\newblock URL \url{https://doi.org/10.1371/journal.pcbi.1010100}.
\newblock Publisher: Public Library of Science.

\bibitem[LeJeune et~al.(2024)LeJeune, Ghaffarzadegan, Childs, and
  Saucedo]{lejeune_mathematical_2024}
Leah LeJeune, Navid Ghaffarzadegan, Lauren~M. Childs, and Omar Saucedo.
\newblock Mathematical analysis of simple behavioral epidemic models.
\newblock \emph{Mathematical Biosciences}, 375:\penalty0 109250, 2024.
\newblock ISSN 0025-5564.
\newblock \doi{https://doi.org/10.1016/j.mbs.2024.109250}.
\newblock URL
  \url{https://www.sciencedirect.com/science/article/pii/S002555642400110X}.

\bibitem[LeJeune et~al.(2025)LeJeune, Ghaffarzadegan, Childs, and
  Saucedo]{lejeune_formulating_2025}
Leah LeJeune, Navid Ghaffarzadegan, Lauren~M. Childs, and Omar Saucedo.
\newblock Formulating human risk response in epidemic models: {Exogenous} vs
  endogenous approaches.
\newblock \emph{European Journal of Operational Research}, 324\penalty0
  (1):\penalty0 246--258, 2025.
\newblock ISSN 0377-2217.
\newblock \doi{https://doi.org/10.1016/j.ejor.2025.01.004}.
\newblock URL
  \url{https://www.sciencedirect.com/science/article/pii/S0377221725000049}.

\bibitem[Bayer(2004)]{bayer2004business}
Steffen Bayer.
\newblock Business dynamics: Systems thinking and modeling for a complex world.
\newblock \emph{Interfaces}, 34\penalty0 (4):\penalty0 324--326, 2004.
\newblock ISSN 00922102, 1526551X.
\newblock URL \url{http://www.jstor.org/stable/25062926}.

\bibitem[Giabbanelli(2023)]{giabbanelli2023gpt}
Philippe~J. Giabbanelli.
\newblock Gpt-based models meet simulation: How to efficiently use large-scale
  pre-trained language models across simulation tasks.
\newblock In \emph{2023 Winter Simulation Conference (WSC)}, pages 2920--2931,
  2023.
\newblock \doi{10.1109/WSC60868.2023.10408017}.

\bibitem[Xu et~al.(2021)Xu, Liu, Cao, Huang, Liu, Qian, Liu, Wu, Dong, Qiu,
  Qiu, Hua, Su, Wu, Xu, Han, Fu, Yin, Liu, Roepman, Dietmann, Virta, Kengara,
  Zhang, Zhang, Zhao, Dai, Yang, Lan, Luo, Liu, An, Zhang, He, Cong, Liu,
  Zhang, Lewis, Tiedje, Wang, An, Wang, Zhang, Huang, Lu, Cai, Wang, and
  Zhang]{xu_artificial_2021}
Yongjun Xu, Xin Liu, Xin Cao, Changping Huang, Enke Liu, Sen Qian, Xingchen
  Liu, Yanjun Wu, Fengliang Dong, Cheng-Wei Qiu, Junjun Qiu, Keqin Hua, Wentao
  Su, Jian Wu, Huiyu Xu, Yong Han, Chenguang Fu, Zhigang Yin, Miao Liu, Ronald
  Roepman, Sabine Dietmann, Marko Virta, sv~Fredrick Kengara, Ze~Zhang, Lifu
  Zhang, Taolan Zhao, Ji~Dai, Jialiang Yang, Liang Lan, Ming Luo, Zhaofeng Liu,
  Tao An, Bin Zhang, Xiao He, Shan Cong, Xiaohong Liu, Wei Zhang, James~P.
  Lewis, James~M. Tiedje, Qi~Wang, Zhulin An, Fei Wang, Libo Zhang, Tao Huang,
  Chuan Lu, Zhipeng Cai, Fang Wang, and Jiabao Zhang.
\newblock Artificial intelligence: {A} powerful paradigm for scientific
  research.
\newblock \emph{The Innovation}, 2\penalty0 (4):\penalty0 100179, 2021.
\newblock ISSN 2666-6758.
\newblock \doi{https://doi.org/10.1016/j.xinn.2021.100179}.
\newblock URL
  \url{https://www.sciencedirect.com/science/article/pii/S2666675821001041}.

\bibitem[Boiko et~al.(2023)Boiko, MacKnight, Kline, and
  Gomes]{Boiko2023AutonomousChemicalResearch}
Daniil~A. Boiko, Robert MacKnight, Ben Kline, and Gabe Gomes.
\newblock Autonomous chemical research with large language models.
\newblock \emph{Nature}, 624:\penalty0 570--578, 2023.
\newblock \doi{10.1038/s41586-023-06792-0}.

\bibitem[Bushaj et~al.(2023)Bushaj, Yin, Beqiri, Andrews, and
  B{\"u}y{\"u}ktahtak{\i}n]{bushaj_simulation-deep_2023}
Sabah Bushaj, Xuecheng Yin, Arjeta Beqiri, Donald Andrews, and {\.I}.~Esra
  B{\"u}y{\"u}ktahtak{\i}n.
\newblock A simulation-deep reinforcement learning ({SiRL}) approach for
  epidemic control optimization.
\newblock \emph{Annals of Operations Research}, 328\penalty0 (1):\penalty0
  245--277, September 2023.
\newblock ISSN 1572-9338.
\newblock \doi{10.1007/s10479-022-04926-7}.
\newblock URL \url{https://doi.org/10.1007/s10479-022-04926-7}.

\bibitem[Simon(1955)]{simon1955behavioral}
Herbert~A. Simon.
\newblock A behavioral model of rational choice.
\newblock \emph{The Quarterly Journal of Economics}, 69\penalty0 (1):\penalty0
  99--118, 1955.
\newblock ISSN 00335533, 15314650.
\newblock URL \url{http://www.jstor.org/stable/1884852}.

\bibitem[Simon(1991)]{simon1991bounded}
Herbert~A. Simon.
\newblock Bounded rationality and organizational learning.
\newblock \emph{Organization Science}, 2\penalty0 (1):\penalty0 125--134, 1991.
\newblock ISSN 10477039, 15265455.
\newblock URL \url{http://www.jstor.org/stable/2634943}.

\bibitem[Li and Wu(2025)]{li2025let}
Zewei Li and Qi~Wu.
\newblock Let it go or control it all? the dilemma of prompt engineering in
  generative agent-based models.
\newblock \emph{System Dynamics Review}, 41\penalty0 (3):\penalty0 e70008,
  2025.
\newblock \doi{https://doi.org/10.1002/sdr.70008}.
\newblock URL \url{https://onlinelibrary.wiley.com/doi/abs/10.1002/sdr.70008}.
\newblock e70008 6647586.

\end{thebibliography}

\end{document}